\documentclass[prd,twocolumn,amsmath,amssymb,floatfix,superscriptaddress,nofootinbib]{revtex4-1}
\usepackage{bm}
\usepackage{amsmath}
\usepackage{epsfig}
\usepackage{xcolor}
\usepackage{natbib}
\usepackage{textcase}
\usepackage{graphicx}
\usepackage{ifthen}
\usepackage{xstring}
\usepackage{bbm}
\usepackage{graphicx}
\usepackage[utf8]{inputenc} 
\usepackage{amssymb}
\usepackage{csquotes}
\usepackage{latexsym}
\usepackage{epstopdf}
\usepackage{comment}
\usepackage{fontawesome}

\usepackage{aas_macros}
\usepackage{xargs}
\renewcommandx{\citet}[3][1={},2={}]{Ref.~\cite[#1][#2]{#3}}

\DeclareGraphicsExtensions{.ps, .png}
\usepackage{dcolumn} 
\usepackage{multirow}
\usepackage{appendix}
\usepackage{footnote}

\usepackage{tabularx,ragged2e,booktabs}
\usepackage[normalem]{ulem}
\usepackage{float}
\restylefloat{table}

\definecolor{RedWine}{rgb}{0.743,0,0}
\definecolor{GrassGreen}{rgb}{0.125,0.75,0.125}
\definecolor{RoyalBlue}{rgb}{0.25,0.41,0.88}
\definecolor{darkgreen}{cmyk}{0.85,0.2,1.00,0.2} 
\definecolor{purple}{cmyk}{0.5,1.0,0,0} 
\definecolor{ultramarine}{rgb}{0.07, 0.04, 0.56}
\definecolor{cadmiumgreen}{rgb}{0.0, 0.42, 0.24}
\definecolor{indigo(dye)}{rgb}{0.0, 0.25, 0.42}

\newcommand{\Mpch}{\mbox{Mpc} \, h^{-1}}
\newcommand{\iMpch}{h \mbox{Mpc}^{-1}}
\newcommand{\fnl}{f_{\rm NL}}
\newcommand{\bhat}[1]{\mathbf{\hat{#1}}}

\newcommand{\fsky}{f_{\rm sky}}

\newcommand{\reffig}[1]{Fig.~\ref{fig:#1}}

\def\max{_{\mathrm{\text{max}}}}
\def\lsim{\mathrel{\raise.3ex\hbox{$$<$$\kern-.75em\lower1ex\hbox{$\sim$}}}}
\def\gsim{\mathrel{\raise.3ex\hbox{$$>$$\kern-.75em\lower1ex\hbox{$\sim$}}}}

\newcommand{\beq}{\begin{equation}}
\newcommand{\eeq}{\end{equation}}

\newcommand{\bea}{\begin{eqnarray}}
\newcommand{\eea}{\end{eqnarray}}

\def\ba#1\ea{\begin{align}#1\end{align}}

\newcommand{\vk}{\mathbf{k}}

\newcommand{\vq}{\mathbf{q}}
\renewcommand{\vr}{\mathbf{r}} 
\newcommand{\vs}{\nonumber\\}
\newcommand{\vvs}{\mathbf{s}}

\newcommand{\vx}{\mathbf{x}}

\newcommand{\khat}{\hat{\vk}}
\newcommand{\qhat}{\hat{\vq}}
\newcommand{\rhat}{\hat{\vr}}

\newcommand{\dd}{\mathrm{d}}
\newcommand{\sums}[2][]{\sum^{\substack{#1}}_{\substack{#2}}}

\newcommand{\tj}[6]{ \begin{pmatrix}
   #1 & #2 & #3 \\
   #4 & #5 & #6 
  \end{pmatrix}}

\def\mnras{Mon.\ Not.\ R.\ Astron.\ Soc.\ }
\def\physrep{Phys.~Rep.}

\usepackage[linktocpage=true]{hyperref}
\hypersetup{
colorlinks=true,
citecolor=ultramarine,
linkcolor=blue,
urlcolor=blue,
pdfauthor={},
pdftitle={},
pdfsubject={}
}
\usepackage[capitalise]{cleveref}
\creflabelformat{equation}{#2\textup{#1}#3}

\begin{document}
	
\title{Wide-Angle Effects in the Power Spectrum Multipoles in Next-Generation Redshift Surveys}

\author{Joshua N. Benabou}
\email{joshua\_benabou@berkeley.edu}
\affiliation{Berkeley Center for Theoretical Physics, University of California, Berkeley, CA 94720, U.S.A.}
\affiliation{Theoretical Physics Group, Lawrence Berkeley National Laboratory, Berkeley, CA 94720, U.S.A.}

\author{Isabel Sands}
\email{isabel\_sands@caltech.edu}
\affiliation{Jet Propulsion Laboratory, California Institute of Technology, Pasadena, California 91109, USA}

\author{Henry S. Grasshorn Gebhardt}
\affiliation{California Institute of Technology, 1200 E. California Boulevard, Pasadena, CA 91125, USA}
\affiliation{Jet Propulsion Laboratory, California Institute of Technology, Pasadena, California 91109, USA}

\author{Chen Heinrich}
\affiliation{California Institute of Technology, 1200 E. California Boulevard, Pasadena, CA 91125, USA}

\author{Olivier Dor\'e}
\affiliation{California Institute of Technology, 1200 E. California Boulevard, Pasadena, CA 91125, USA}
\affiliation{Jet Propulsion Laboratory, California Institute of Technology, Pasadena, California 91109, USA}

\begin{abstract}

As galaxy redshift surveys expand to larger areas on the sky, effects coming from the curved nature of the sky become important, introducing wide-angle (WA) corrections to the power spectrum multipoles at large galaxy-pair separations. These corrections particularly impact the measurement of physical effects that are predominantly detected on large scales, such the local primordial non-Gaussianities. In this paper, we examine the validity of the perturbative approach to modeling WA effects in the power spectrum multipoles for upcoming surveys by comparing to measurements on simulated galaxy catalogs using the Yamamoto estimator. We show that on the scales $k \lesssim 2\pi/\chi$, where $\chi$ is the comoving distance to the galaxies,
the estimated power spectrum monopole differs by up to $5\%$ from the second-order perturbative result, with similar absolute deviations for higher multipoles. To enable precision comparison, we pioneer an improved treatment of the $\mu$-leakage effects in the Yamamoto estimator. Additionally, we devise a solution to include $f_{\rm NL}$ in the perturbative WA calculations, avoiding divergences in the original framework through the integral constraint. This allows us to conclude that WA effects can mimic 
a $f_{\rm NL}\sim5$ signal in the lowest SPHEREx redshift bin. We recommend using non-perturbative methods to model large scale power spectrum multipoles for $f_{\rm NL}$ measurements. A companion paper, Wen et al. 2024, addresses this by introducing a new non-perturbative method going through the spherical Fourier-Bessel basis. 

\end{abstract}

\maketitle

\section{Introduction}
\label{sec:intro}

Current and upcoming galaxy redshift surveys such as
DESI\footnote{\url{http://desi.lbl.gov/}}~\cite{Aghamousa:2016zmz}, 
SPHEREx\footnote{\url{https://spherex.caltech.edu/}}~\cite{Dore:2014cca},
\emph{Euclid}\footnote{\url{https://www.euclid-ec.org}}~\cite{2011arXiv1110.3193L}
and 
the \textit{Nancy Grace Roman} Space Telescope\footnote{Formerly the Wide Field Infrared Survey Telescope (WFIRST): \url{https://roman.gsfc.nasa.gov/}} \cite{Spergel:2013tha}
will measure the galaxy distribution over an increasing volume, probing interesting physical effects that become important on large physical scales. One such example is measuring the local primordial non-Gaussianity $\fnl^{\rm local}$ to $\mathcal{O}(1)$ precision with SPHEREx, which can help us to distinguish between multi-field and single-field models of inflation~\cite{Dore:2014cca}. 
However, measuring the ultra-large scales comes with new challenges: The curved nature of the sky also becomes more apparent on large angular scales, and any effects that depend on the line-of-sight (LOS) of the observer need to be modeled accurately to compute $n$-point statistics. 

For example, the redshift space distortions (RSD) effects depend on the LOS of individual galaxies. In small-area surveys, the global plane-parallel approximation, in which all galaxies are assumed to have the same LOS, suffices; in  surveys where the galaxy separations become sufficiently large, wide-angle corrections must be included to properly account for the effects due to the different LOS. Traditionally, these corrections have been introduced either perturbatively as an expansion around the plane-parallel approximation, or non-perturbatively, by computing the correlation function and transforming back into Fourier space \cite{Castorina+:2022JCAP...01..061C}. While the perturbative expansion seems desirable in terms of its potential computational simplicity, the non-perturbative expansion is expected to be fully accurate, as it makes no assumption about the smallness of the small parameter, the galaxy angular separation. For example, as we will discuss, the wide-angle expansion breaks down when the separation between the galaxies exceeds the LOS distance to the pair. For thin redshift bins for which galaxies within the bin can be approximated as equidistant to the observer, this corresponds to angular separations larger than $60^\circ$.

In the case of measuring $\fnl^{\rm local}$ (we will drop the superscript ``local" from now on), this leads naturally to the following questions. How important are those wide-angle effects for constraining $\fnl$? 
Is the perturbative method accurate enough for modeling them for the purpose of constraining $\fnl$ with SPHEREx?
If so, do we need to include terms beyond second-order, which was sufficient for the BOSS analysis in Ref.~\cite{Beutler+:2019JCAP...03..040B}? In this paper, we set out to answer these questions by enabling the perturbative WA calculation of the galaxy power spectrum including $\fnl$ effects, and by comparing the WA modeling to simulations.

We begin by developing a technique to include $\fnl$ in the perturbative WA modeling framework developed in Beutler et~al.~\cite{Beutler+:2019JCAP...03..040B}, through the integral constraint. We apply it to show that WA effects mimic 
an $\fnl\sim 5$ signal, confirming the importance of modeling them for the SPHEREx survey. 
We then compare our modeling with simulations, in order to determine whether the perturbative expansion up to order $n = 2$ would be sufficient. To do so, we refine the traditional implementation of the Yamamoto estimator to accurately handle the $\mu$-leakage -- an effect that comes from averaging over a discrete number of Fourier cells within each $k$-shell. We also find that we need to include in our 
modeling the $\alpha$ term -- the Newtonian Doppler term that vanishes in the plane-parallel limit.

After careful work to eliminate discrepancies between the estimator implementation and the modeling of its expectation value, we find a remaining discrepancy, concluding that the $n = 2$ perturbative expansion is not enough. Furthermore, we find large corrections when including $n=4$ terms, which are also insufficient to explain the discrepancy. In a companion paper, we propose a new method to obtain the non-perturbative model, and show that it agrees well with the same set of simulations used here \cite{Wen+:2024}. This new method goes through the spherical Fourier-Bessel space in order to capture all LOS effects fully non-perturbatively, before transforming back into the power spectrum multipole space (c.f. with the existing non-perturbative method on the market through the correlation function). 

Note that the plane-parallel approximation suffices for surveys with small areas, while the perturbative model is sufficient for large-area surveys with a particular scale cut that depends on the redshift considered. So non-perturbative WA modeling is not necessary for all future surveys, but is relevant for those whose science cases depend on the ultra-large scales, such as measuring $\fnl$ with SPHEREx, or constraining general relativistic (GR) effects which are LOS effects that also become important on similar scales. 

This paper is structured as follows. In section~\ref{sec:background}, we describe the theory of wide-angle RSD effects according to the formalism from Ref.~\cite{Beutler+:2019JCAP...03..040B}.
In section~\ref{sec:results}, we detail our methodology for computing wide-angle effects and discuss results for BOSS, Roman, Euclid, and SPHEREx, including $\fnl$ in our calculations. In \cref{sec:simulations} we explore the WA effect with log-normal simulations. In section~\ref{sec:comparison}, we compare our analytic results to measurements of the power spectrum multipoles from simulated galaxy catalogs. Lastly, we conclude and discuss future work in section \ref{sec:conclusion}. The wide-angle formulae for the correlation function multipoles along with other technical details of our simulations and power spectrum estimator are given in the appendices.

\section{Background}
\label{sec:background}

The two-point correlation function of the galaxy density contrast $\delta_g$ is defined as
\beq
\xi(\mathbf{s_1},\mathbf{s_2})\equiv\langle \delta_g(\mathbf{s_1})\delta_g(\mathbf{s_2}) \rangle \,,
\label{eq:correlation_function_general}
\eeq
where the brackets denote an ensemble average, and $\vvs_1$ and $\vvs_2$ are
positions in redshift space.
The galaxy velocity field in Fourier space is, to first-order in the linear matter density field, 
\begin{equation}
\textbf{v(k)}
= i \, a H(a) \, f(a) \, \frac{\textbf{k}}{k^2} \, \delta_m(\textbf{k}) \,,
\label{eq:vk}
\end{equation} 
where $a$ is the scale factor, $H$ is the Hubble parameter, $f$ the logarithmic
growth factor, and $\delta_m$ is the matter density contrast. The peculiar
velocity of the galaxies induces a shift in their observed position
$\mathbf{s}_i$ via
\begin{equation}
    \textbf{s}_i
    =
    \textbf{r}_i
    + \frac{\big(\textbf{v}_i\cdot \bhat{d}\big)\,\bhat{d}}{a H(a)} \,,
    \label{eq:doppler_shift}
\end{equation}
where
$\textbf{r}_i$ is the comoving position vector in real space, and $\bhat{d}$ is the LOS unit vector. For an observer at the origin, $\bhat d=\bhat r_i$. Hence, redshift-space distortions (RSD) break the statistical homogeneity and isotropy of the galaxy correlation,
i.e., we cannot express it as $\xi(\mathbf{s_1},\mathbf{s_2}) = \xi(s)$, with $s=|\bf{s}|=|\mathbf{s_2}-\mathbf{s_1}|$ the distance separating the galaxy pair. However, in the global plane-parallel approximation, where a common LOS for all galaxies is assumed, we may still write $\xi(s, \mu)$ as a function of the galaxy separation distance $s$ and $\mu=\bhat{s} \cdot \bhat{d}$, the cosine of the angle between the separation vector $\bhat{s}$ and the LOS vector $\bhat{d}$.

As surveys grow in angular extent, the global plane-parallel approximation starts to break down. Refs.~\cite{Castorina_2018, Reimberg_2016} developed a formalism using the local plane-parallel approximation, where a unique LOS is assigned to each galaxy pair. 
In this case, the correlation function $\xi(s,d,\mu)$ is dependent on one additional parameter, the LOS distance $d$. There are multiple conventions for choosing $\mathbf{d}$ given a galaxy pair: (i) The mean LOS convention where we take  $\mathbf{d}=\frac{1}{2}(\mathbf{s_1}+\mathbf{s_2})$; (ii) the angular bisector convention; (iii) the end-point convention in which $\mathbf{d}$ coincides with one of the galaxy LOS's, e.g., $\mathbf{d}=\mathbf{s_1}$. The mean and angular bisector LOS have the advantage of being symmetric under the exchange of $\mathbf{s_1}$ and $\mathbf{s_2}$, but require an $O(N^2)$ computation for the correlation function (but see \cite{Philcox:2021tfv}). 
In the end-point case, the power spectrum estimator is an integral which can be separated into products of lower-dimensional integrals that may be evaluated via a fast-Fourier-transform (FFT)~\cite{Hand17_FFT}, and therefore becomes more computationally tractable.
This is the convention adopted in practice and that we will use throughout this work.

\subsection{Modeling of wide-angle effects}
\label{sec:theory}

Following Ref.~\cite{Beutler+:2019JCAP...03..040B}, we decompose the correlation function in a basis of Legendre polynomials and further expand the multipoles in the parameter $x_s=s/d$:
\ba
\xi(s,d,\mu) = \sum_l \xi_l(s,d)\mathcal{L}_l(\mu) = \sum_{l,n} x_s^n \xi_l^{(n)}(s)\mathcal{L}_l(\mu).
\label{eq:correlation_multipoles_expansion}
\ea
In the global plane-parallel approximation where
$x_s \to 0$, only the $n = 0$ term survives, and we recover the classic Kaiser result 
where the only nonzero $\xi_l^{(n)}$ are $l=0,2,4$~\cite{original_kaiser}:
\ba
\xi_0^{(0)} (s)   &=    \left( 1 + \frac{2}{3}  \beta +  \frac{1}{5} \beta^2 \right)b_1^2 \Xi_0^{0}(s)\,, \label{eq:kaiser_0}\\
\xi_2^{(0)} (s)   &=  -   \left(\frac{4}{3} \beta + \frac{4}{7} \beta^2 \right)b_1^2 \Xi_2^{0}(s)\,, \label{eq:kaiser_2}\\
\xi_4^{(0)} (s)   &=    \frac{8}{35}  \beta^2b_1^2 \Xi_4^{0}(s)\,, \ \label{eq:kaiser_4}
\ea
where $b_1$ is the linear galaxy bias, $\beta=f/b_1$ is the RSD parameter, and $f = \mathrm{d}\ln D/\mathrm{d}\ln a$ is the logarithmic growth rate. We also denote the Hankel transform of the real-space matter power spectrum $P(k)$ by
\begin{equation}
  \Xi_\ell^{(n)}(s) \equiv \int\frac{k^2\,\mathrm{d}k}{2\pi^2}\ (k)^{-n}P(k)\,j_\ell(ks) \,,
  \label{eq:Hankel_transform}
\end{equation}
with $j_\ell$ the spherical Bessel function of order $\ell$. For further details on the calculation of the coefficients $\xi^{(n)}_l(s)$, see Ref.~\cite{Reimberg_2016}.

Crucially, the expansion Eq.~\ref{eq:correlation_multipoles_expansion} is only valid for $x_s<1$. Thus, for example, for a pair of galaxies which are equidistant to the observer (note that all pairs of galaxies approach this limit for a sufficiently thin redshift bin), one should expect the wide-angle expansion to be inappropriate for modes corresponding to angular separations between the pair exceeding $60^\circ$. For smaller opening angles, the expansion is still invalid for pairs whose separation distance $s$ is larger than the LOS distance $d$.
To understand the regime of validity of the expansion, note that it is obtained by expressing the operators which map the Fourier space density field $\delta(\mathbf{k})$ to the configuration space densities $\delta(\mathbf{s_1})$ and $\delta(\mathbf{s_2})$  in terms of the distances $s_1, s_2$ to the galaxies and the angle $\hat{\mathbf{s}}_1\cdot\hat{\mathbf{s}}_2$. From the law of cosines, we can write
$\hat{\mathbf{s}}_1\cdot\hat{\mathbf{s}}_2=(1-x_s  \mu) /\sqrt{1 -2x_s\mu+x_s^2}$
and $s_2=s_1 \sqrt{1-2 x_s \mu+x_s^2}$. Following the derivation of Ref.~\cite{Reimberg_2016} (their Eq.~4.15), the correlation function multipoles are then calculated by Taylor expanding these operators in $x_s$, about $x_s=0$. This amounts to expanding a rational function of $x_s$ with denominator $1-2 x_s \mu+x_s^2$. 
The Taylor series of $(1-2 x_s \mu+x_s^2)^{-1}$ has a radius of convergence equal to unity, independently of $\mu$ \footnote{Taylor expanding gives $(1-2 x_s \mu+x_s^2)^{-1}=\sum_n{ U_n(\mu)x_s^{n}}$ where $U_n$ is the Chebyshev polynomial of the second kind \cite{NIST:DLMF}. Writing $\mu = \cos\theta$, one has $U_n(\mu)=\sin((n+1)\theta)/\sin(\theta)$, such that the radius of convergence of this series is unity from the Cauchy-Hadamard Theorem.}. 

For nearly full-sky surveys such as SPHEREx, we are no longer in the regime of validity of the perturbative analysis for a large portion of galaxy pairs in the sample (see Appendix \ref{sec:counting_galaxy_pairs} for an estimate of this fraction as a function of the separation distance $s$).
However, further investigation is necessary to establish (i) which order $n$ is sufficient on scales for which $x_s<1$, and (ii) whether the expansion remains an acceptable approximation (at any order) beyond this threshold, in particular for a $f_\mathrm{NL}$ measurement at order unity precision. Indeed, in this work we explore both of these questions using simulated SPHEREx-like galaxy catalogs,
and find that the perturbative formalism is insufficient in the latter case. To address (i) we compute wide-angle corrections to order $n=4$.

In the bisector and mean LOS, the symmetry of the galaxy pair about the LOS and the parity of the Legendre polynomials imply that $\xi_l^{(n)}=0$ for odd $l$ and $n$. The end-point LOS however breaks this symmetry, giving rise to odd multipoles at order $n=1$. These odd multipoles are of geometric origin and do not contain any cosmological information, but may leak into measurements of the physical dipole generated by relativistic effects \cite{Raccanelli_2014}. We give the higher order $\xi_l^{(n)}$ for $n \le 4$ in Appendix \ref{app:wide_angle_corrections}, and to aid reproducibility we provide a \texttt{Mathematica} notebook 
\href{https://github.com/joshua-benabou/wide_angle_expansion
}
{\faGithub}, 
which computes the $\xi^{(n)}_l$ symbolically for arbitrary $(l,n)$ \footnote{We caution that there are discrepancies in the expressions of the $\xi_l^{(n)}$ found in literature. In particular Refs. \cite{Beutler+:2019JCAP...03..040B} and \cite{Castorina_2018} give  expressions for the $n=2$ correlation function multipoles in the bisector and endpoint LOS conventions which are inconsistent with those in Ref.~\cite{Reimberg_2016}. The notebook generates terms in agreement with Ref.~\cite{Reimberg_2016}. 
}.

A fully consistent treatment of the wide-angle corrections should account for the galaxy selection function. Indeed, the real-space density contrast is mapped into redshift space via
\ba
\delta_s(\mathbf{s})=\delta(\mathbf{r})-\left(\frac{\partial}{\partial r}+\frac{\alpha}{r}\right)\frac{\mathbf{v}\cdot\hat{\mathbf{r}}}{aH}\,,
\ea 
with
\ba
\alpha\equiv\frac{\partial\left[\ln\!\left(\bar{n}(\mathbf{r}) \, r^2\right)\right]}{\partial \ln r} \,,
\label{eq:alpha}
\ea
where $\bar{n}(\mathbf{r})$ is the
expected galaxy number density at position $\vr$, in units of inverse comoving volume.
The $\alpha/r$ term encodes the selection function. In realistic surveys, selection function modeling is only possible if the real space galaxy density is known, though this is usually a difficult undertaking.
Indeed, for this reason and the fact that it is often a small effect, the authors of Ref.~\cite{Beutler+:2019JCAP...03..040B} did not account for it when comparing the wide-angle expansion (up to $n=2$) to BOSS DR12 data. On the other hand, in our simulations we use a constant galaxy number density, such that $\alpha=2$ above, and we find that in this case these terms are non-negligible on scales where the perturbative expansion is valid.

The ensemble average of the power spectrum estimator in the endpoint
convention (see details of our implementation in
\cref{sec:Yamamoto_estimator}), convolved with the survey window
$W(\mathbf{s})$, is given by Ref.~\cite{Beutler+:2019JCAP...03..040B}

\begin{widetext}
\ba
{P}_{l}(k)
&=
\frac{2l+1}{V_\mathrm{survey}}
\int \frac{\mathrm{d} \Omega_k}{ 4\pi} \, \mathrm{d}^3 s_1\,\mathrm{d}^3 s_2\, e^{i \mathbf{k} \cdot \mathbf{s}}\,\langle\delta(\mathbf{s_1})\delta(\mathbf{s_2})\rangle\, \,W(\mathbf{s_1})W(\mathbf{s_2})\,\mathcal{L}_{l}(\hat{\mathbf{k}}\cdot\hat{\mathbf{s}}_1) 
\label{eq:power_estimator}
\\
&=
(-i)^{l} (2l+1) \sum_{l',
L}\begin{pmatrix}
l' & L & l \\
0 & 0 & 0
\end{pmatrix}^2\int \mathrm{d}s\, j_{l} (ks) \sum_n s^{n+2} \,\xi_{l'}^{(n)}(s) Q_L^{(n)}(s) \,,
\label{eq:convpower} 
\ea
\end{widetext}
where $\begin{pmatrix}
l' & L & l\\
0 & 0 & 0
\end{pmatrix}$ is the Wigner-3j symbol.
We denote the window multipole at order $n$ in $x_s$ by \cite{Beutler+:2019JCAP...03..040B}
\ba
Q_L^{(n)}(s)
&\equiv
4\pi \, (2L+1)\int \frac{\dd\Omega_s}{4\pi}
\int \frac{\dd^3d}{V_\mathrm{survey}} \, d^{-n}
\nonumber\\&\quad\times
W(\mathbf{s_1}(\mathbf{s},\mathbf{d}))
\,W(\mathbf{s_2}(\mathbf{s},\mathbf{d}))\,\mathcal{L}_L(\hat{\mathbf{s}}\cdot\hat{\mathbf{d}})\,. 
\label{eq:Q_window_multipole}
\ea
For the window function we use the normalization $V_\mathrm{survey}^{-1}\int d^3s\, W(\mathbf{s})=1$, where the window vanishes outside the volume of interest, delimited by the angular footprint of the survey and radially by the given redshift bin, and we normalize the window multipoles by the survey volume $V_\mathrm{survey}$. Eq.~\ref{eq:convpower} indicates that the wide-angle effects couple the window multipoles (which carry purely geometric information) to the correlation function multipoles (which quantify RSD).

\section{Wide-angle modeling results}
\label{sec:results}

We now apply the perturbative formalism to model wide-angle corrections to the power spectrum multipoles for a selection of next-generation surveys and discuss how primordial non-Gaussianity (PNG) measurements are biased by wide-angle effects. In the following sections we fiducially assume a flat Planck 2018 cosmology ~\cite{Aghanim:2018eyx}. 
We calculate the real-space linear matter power spectrum and linear growth factor with the  
\texttt{CAMB}\footnote{\texttt{CAMB}: \url{https://camb.info/}} package~\cite{camb}. The remainder of the calculation\footnote{Our code, 
which computes the wide-angle expansion for the power spectrum multipoles for a given survey window and choice of LOS (mean, bisector, or endpoint), will be made available upon request.} is in Julia \cite{Julia-2017}. 

\subsection{Window multipoles}
\label{sec:window_multipoles}

\begin{figure}
\includegraphics[width=0.45\textwidth]{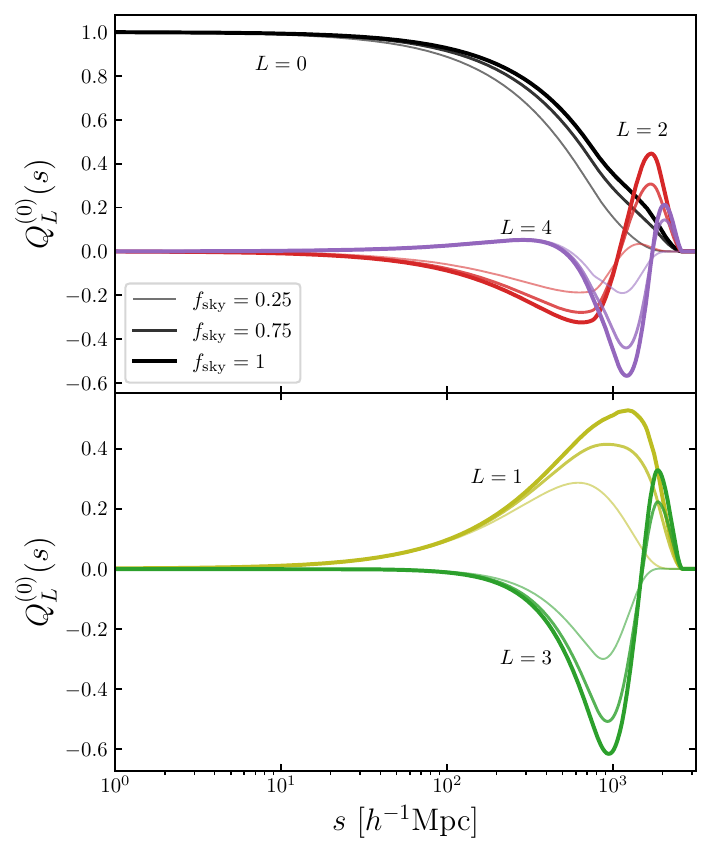}
\caption{Window multipoles $L \le 4$ for synthetic masks of sky coverages $f_{\text{sky}}=25\%, \,75\%, \, 100\%$ for the redshift bin $[0.2, 0.5]$, corresponding to the comoving radial-distance bin  $[571,1317] \, \iMpch$. For this plot only we use the normalization $Q_0^{(0)}(s \to 0)=1$.
Curves corresponding to the same survey mask are distinguished by line thickness and opacity which both increase with sky coverage.
}
\label{fig:window_multipoles_synthetic_masks_comparison}
\end{figure}

To obtain the convolved power spectrum multipoles $P_{l}(k)$ in Eq.~\ref{eq:convpower} up to $l = 4$, we need to first calculate the correlation function multipoles $\xi_{l'}^{(n)}$, which have non-vanishing terms up to $l'=6$ ($l'=8$) if working to order $n=2$ ($n=4$)\footnote{Note that in Ref.~\cite{Beutler+:2019JCAP...03..040B}, measurements of the power spectrum multipoles from BOSS data were compared to the wide-angle expansion at order $n=2$, but only the correlation function multipoles with $l' \le 4$ were included. Ignoring the $l'=5,6$ terms is likely an acceptable approximation within the error bars of BOSS data, however we find doing so leads to $\mathcal{O}(1)$ differences in the hexadecapole. In this work we include all terms for consistency. }. The Wigner-3j symbols then impose that we include the window multipoles up to $L=8$ ($L=12$).
To compute the window multipoles, we assume a separable window function that vanishes outside out the survey footprint and the redshift bin of interest $[z_{\text{min}},z_{\text{max}}]$:
\ba
W(\mathbf{s})
=
\mathbf{1}_{[\chi_{\text{min}},\chi_{\text{max}}]}(s)
\,W(\hat{\mathbf{s}}) \,,
\label{eq:window_function}
\ea
where $\mathbf{1}$ is an indicator function equal to unity inside the interval $[\chi_{\rm min}, \chi_{\rm max}]$, where $\chi$ is the comoving radial distance. 
Here 
$W(\hat{\mathbf{s}})$ is the angular mask and equal to unity inside the survey footprint on the sky.
Thus, the survey volume is $V_{\text{survey}}=\frac{4\pi}{3}(\chi_{\text{max}}^3-\chi_{\text{min}}^3)f_{\text{sky}}$, where
we denote by $f_{\rm sky}$ the fraction of sky covered by the survey footprint. 
In this paper we do not consider more general radial selection functions.

In \reffig{window_multipoles_synthetic_masks_comparison}, we show the window multipoles $Q^{(0)}_L(s)$ 
up to $L = 4$ for a series of circular synthetic masks with $f_{\rm sky} = 0.25$, 0.75 and 1. We use the redshift bin $z \in [0.2,0.5]$, corresponding to a comoving-distance range with $\chi_{\rm min} = 571\, \iMpch$ and $\chi_{\rm max} = 1317\, \iMpch$. For the purpose of illustration only, we further normalize the $Q_L^{(0)}$ with $1/(4\pi)$ such that $Q_0^{(0)}(s \to 0)=1$.
Note that $Q_L^{n}(s)$ vanishes for $L>0$ on the smallest scales $s \to 0$ and on scales beyond the largest scale in the survey $s \ge 2\chi_{\text{max}}$, and the order of magnitude is set by $[\chi(z_{\rm eff})]^{-n}$ for some effective redshift $z_\mathrm{eff}$ inside the survey.

\cref{fig:window_multipoles_synthetic_masks_comparison} shows that the window multipoles have more power at larger scales for surveys with more sky-coverage, contributing to larger wide-angle effects in the power spectrum multipoles as we shall see next.

\subsection{Power spectrum multipoles for different masks}
\label{sec:power spectrum multipoles}

\begin{figure*}
\includegraphics[width=1\textwidth]{"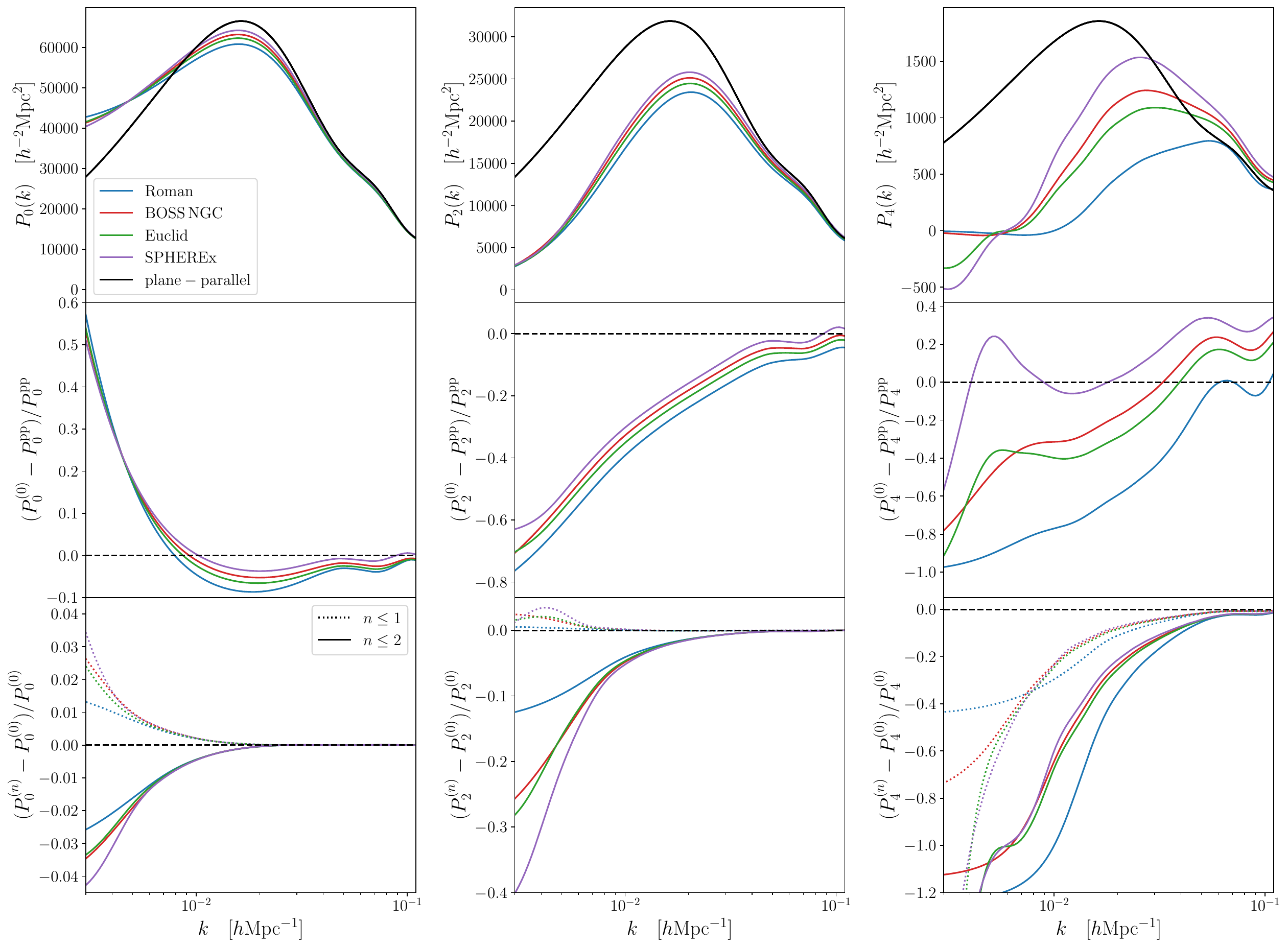"}
\caption{Upper panels: For the monopole, quadrupole, and hexadecapole, comparison between the power spectrum multipoles computed with end-point LOS (to order $n=2$) to the plane-parallel approximation, plotted for \emph{Roman}, BOSS NGC, \emph{Euclid}, and SPHEREx-like masks, for the redshift bin $0.2<z<0.5$ (same redshift range for all masks). Here $z_\mathrm{eff}=0.38$ and the galaxy bias is $b_1=1.8$. Middle panels: Deviation of the convolved spectrum from the global plane-parallel approximation at order $n=0$. Lower panels: Deviation of the convolved power spectrum, at orders $n=1,2$, from $n=0$.
} 
\label{fig:survey_comparison_end-point_vs_kaiser}
\end{figure*}

The power spectrum multipoles are computed from the correlation function multipoles $\xi_l^{(n)}$ in Eq.~\ref{eq:convpower} convolved with the window function multipoles of the previous section. To compute the correlation function multipoles efficiently, we use a variant of the FFTLog method \cite{Grasshorn_Gebhardt_2018} to perform the Hankel transforms of the power spectrum (Eq.~\ref{eq:Hankel_transform})
\footnote{In order for the FFTLog algorithm to yield numerically stable results we extrapolate the power spectrum with power laws to scales $k \gg 1 \, \iMpch$ as well as to scales much larger than the survey size $k \ll 10^{-4} \, \iMpch$.}.

To examine how wide-angle effects vary with the angular survey geometry,
\cref{fig:survey_comparison_end-point_vs_kaiser} compares the multipoles
calculated with several realistic masks corresponding to
\emph{Roman}
($\fsky \simeq
0.06$), BOSS NGC\footnote{\url{https://sdss.org}} ($\fsky \simeq 0.185$),
\emph{Euclid}
($\fsky \simeq 0.36$),
and SPHEREx
($\fsky \simeq 0.8$).\footnote{The angular mask for  
\href{https://irsa.ipac.caltech.edu/data/Planck/release_1/ancillary-data/previews/HFI_Mask_GalPlane_2048_R1.10/index.html}{SPHEREx} and 
\href{https://data.sdss.org/datamodel/files/BOSSTARGET_DIR/data/geometry/boss_survey.html}{BOSS} are hyperlinked; masks for Roman and Euclid were provided by the associated collaborations.}
Each survey has a unique galaxy bias $b_1(z_\mathrm{eff})$, but for simplicity we assume the same constant bias $b_1(z_\mathrm{eff})=1.8$ for all. We also choose the same redshift bin
$[0.2,0.5]$ for all, which actually lies outside of the redshift range of Roman ($0.45$--$2.75$) and Euclid  ($0.7$--$1.8$ for the galaxy clustering sample), but we use Fig.~\ref{fig:survey_comparison_end-point_vs_kaiser} only to illustrate how wide-angle corrections behave as a function of angular geometry.

In the top row of panels in \cref{fig:survey_comparison_end-point_vs_kaiser}, we plot the power spectrum multipoles in the plane-parallel approximation (unconvolved with the window) in black solid, and the convolved power spectrum multipoles for the masks from the aforementioned surveys, including up to $n=2$ terms.
Notable is that the amplitude near the peak is reduced by a greater amount, the smaller the sky fraction $\fsky$ of the mask. 
On very large scales, the windowed power spectra tend to a constant 
(which will no longer be true once the integral constraint/local average effect is included in \cref{sec:nongaussianity}).
This constant is larger for a smaller window, as expected from the larger variance of a smaller volume.

The middle panels show the fractional deviation of the $n = 0$ contribution to the convolved multipoles with respect to the unconvolved one. The $n=0$ term includes the main contribution from the window convolution, but also the leading order wide-angle contribution, loosely speaking going from global to local plane-parallel approximation.
This leading-order term captures the main effects discussed in the previous paragraph.
The hexadecapole is the most severely altered by wide-angle effects and window convolution and shows, already at 
$k \approx 0.1 \, \iMpch$, significant deviations from the plane-parallel approximation.

The bottom panels of Fig.~\ref{fig:survey_comparison_end-point_vs_kaiser} show the fractional deviation of the $n\leq1$ and $n\leq2$ terms compared to the $n=0$ convolved multipoles.
These represent non-local wide-angle terms on top of the $n=0$ local-plane-parallel term.
The $n=0$ term is a good approximation on small scales by construction; 
whereas for a given large scale, surveys covering larger sky fractions would have stronger wide-angle effects. Indeed, it is the angular scales that matter when considering wide-angle effects rather than the physical scales.
Comparing between the different multipoles, we see also that the wide-angle effects start to become important at a smaller scale for higher multipoles. Moreover, at a given (large) scale, these corrections are larger for higher multipoles: at the level of percents for the monopole, tens of percents for the quadrupole, and order unity for the hexadecapole.

The $n=1$ and $n=2$ corrections for the eBOSS NGC (red) and \emph{Euclid}-like (green) masks are similar, despite the sky coverage of \emph{Euclid} being approximately twice that of eBOSS NGC.
This is attributed to the fact that the Euclid survey mask consists of two disjoint patches of roughly $\fsky=0.18$, similar to the BOSS NGC coverage of $\fsky=0.185$.

\subsection{Wide-angle modeling for different redshift bins}
\label{sec:SPHEREx}

\begin{table}
\begin{tabular}{ |p{1.5 cm}||p{1.1 cm}|p{1.3 cm}|p{1.0cm}|}
 \hline
  $z$-bin & $z_\mathrm{eff}$ & $b_1(z_\mathrm{eff})$& $f(z_\mathrm{eff})$\\
 \hline
 $0.2-0.5$  & $0.38$  & 1.5 &  0.715\\
 $0.6-0.8$  & $0.7$   & 1.9 & 0.816 \\
 $1.0-1.6$  & $1.3$   & 2.1  & 0.914 \\
 $2.2-2.8$  & $2.5$   & 4.2 & 0.974 \\
 \hline
\end{tabular}
\caption{Parameters for Fig.~\ref{fig:SPHEREx_effect_of_redshift}.}
\label{table:SPHEREx_effect_of_redshift_table}
\end{table}

\begin{figure*}
\includegraphics[width=1\textwidth]{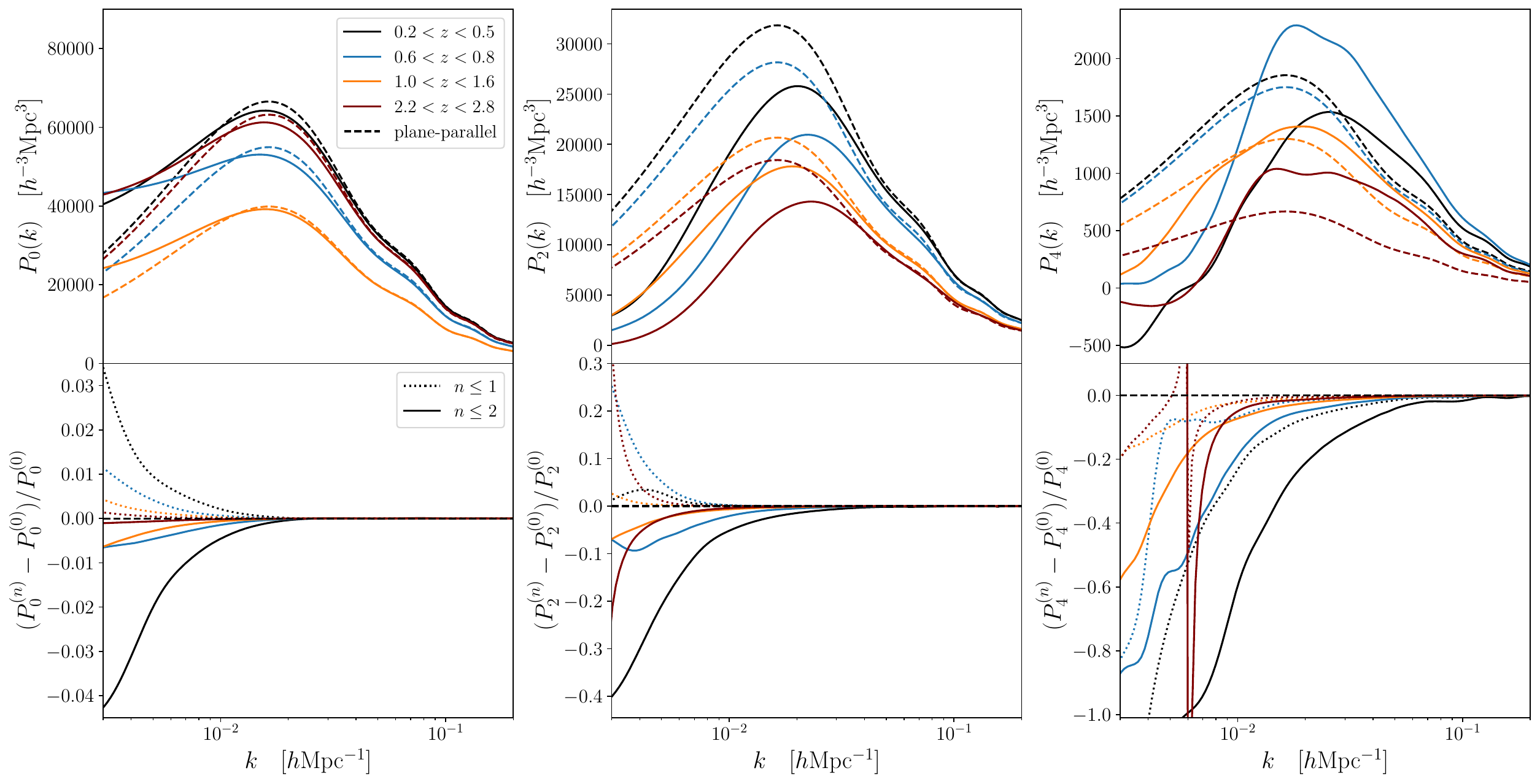}
\caption{Upper panel: Comparison of the convolved power spectrum computed with end-point LOS geometry (at second order in $x_s$) to the plane-parallel approximation, plotted for SPHEREx as a function of redshift bin. The galaxy bias, growth rate, and effective redshifts for this computation are given in table \ref{table:SPHEREx_effect_of_redshift_table}. Lower panel: Deviation of the convolved power spectrum, computed to order $n=1,2$, from order $n=0$.}
\label{fig:SPHEREx_effect_of_redshift}
\end{figure*}

Next we study the wide-angle effects in different redshift bins for SPHEREx. In Fig.~\ref{fig:SPHEREx_effect_of_redshift}, we show the perturbative results specific to SPHEREx for four redshift bins. The galaxy bias, the growth rate and the effective redshifts we use for each bin are given by Table~\ref{table:SPHEREx_effect_of_redshift_table} and obtained from the SPHEREx public GitHub repository\footnote{\url{https://github.com/SPHEREx/Public-products}}. We only show the upper and lower panels of previous figures this time. 

In the upper panel we compare the $n\leq2$ convolved power spectrum (solid) to the plane-parallel approximation (dashed).
The amplitude of the spectra is governed by three distinct effects: the real-space matter power spectrum $P(k,z)$ whose amplitude decreases monotonically with $z$, the bias $b_1(z)$ which we assume increases monotonically with $z$, and the growth rate $f(z)$ which also increases monotonically with $z$. For the bins considered in Fig.~\ref{fig:SPHEREx_effect_of_redshift}, the values of $f$ and $b_1$ are of the same order for the first three bins, and $b_1$ doubles for the fourth bin $2.2<z<2.8$, hence the large increase in power for this bin.

In the lower panel we note that for each multipole the higher order corrections $n=1,2$ to the end-point estimator deviate less strongly from the $n=0$ curve as the redshift increases. Indeed, at lower redshifts the subtended angle at a given scale is larger and wide-angle effects are naturally more important. For 
redshifts $z>0.6$
the $n=1$ and $n=2$ corrections to the monopole are both within $1\%$ of plane-parallel approximation at $k=3 \times 10^{-3} \, \iMpch$. For the quadrupole, however, these higher-order corrections are at the level of $10\%$, even for the highest redshift bin $2.2<z<2.8$.

\begin{figure*}[ht]
\includegraphics[width=1.0\textwidth]{"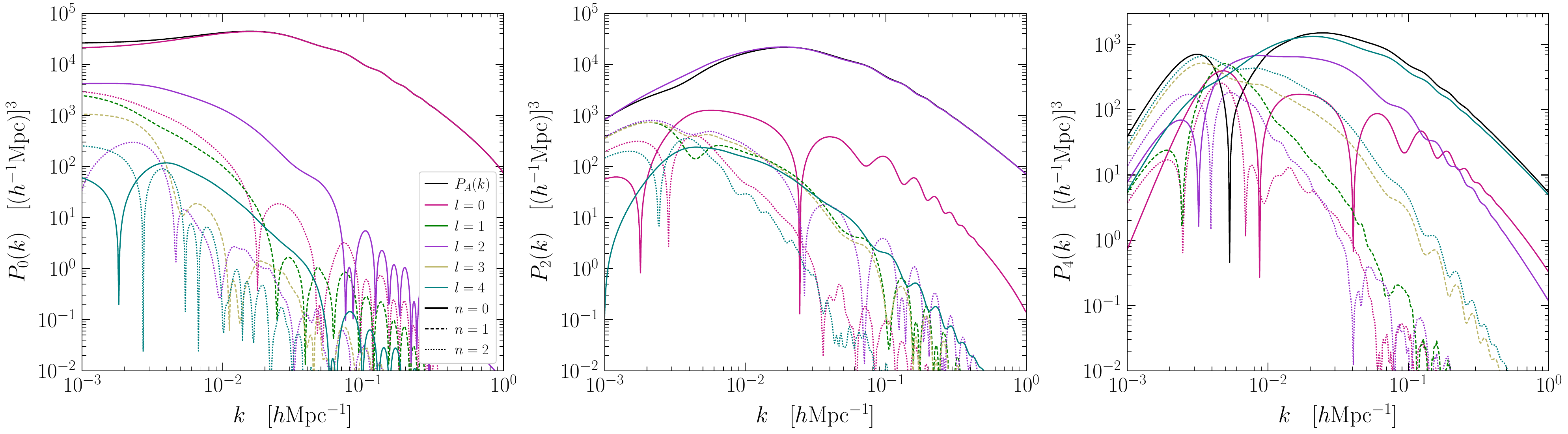"}%
\caption{
The solid black curves show the even multipoles of the convolved power spectrum (end-point estimator) for SPHEREx for the redshift bin 
$0.2<z<0.5$, $z_\mathrm{eff}=0.38$, $b_1=1.455$, $f=0.71$
. Individual contributions from the terms in $\xi_l^{(n)}$ in Eq.~\ref{eq:convpower} are given for $l \le 4$, $n \le 2$.  The $n=0,1,2$ contributions are respectively solid, dashed, dotted.}
\label{fig:SPHEREx_even_multipoles}
\end{figure*}

\begin{figure*}[ht]
\includegraphics[width=0.8\textwidth]{"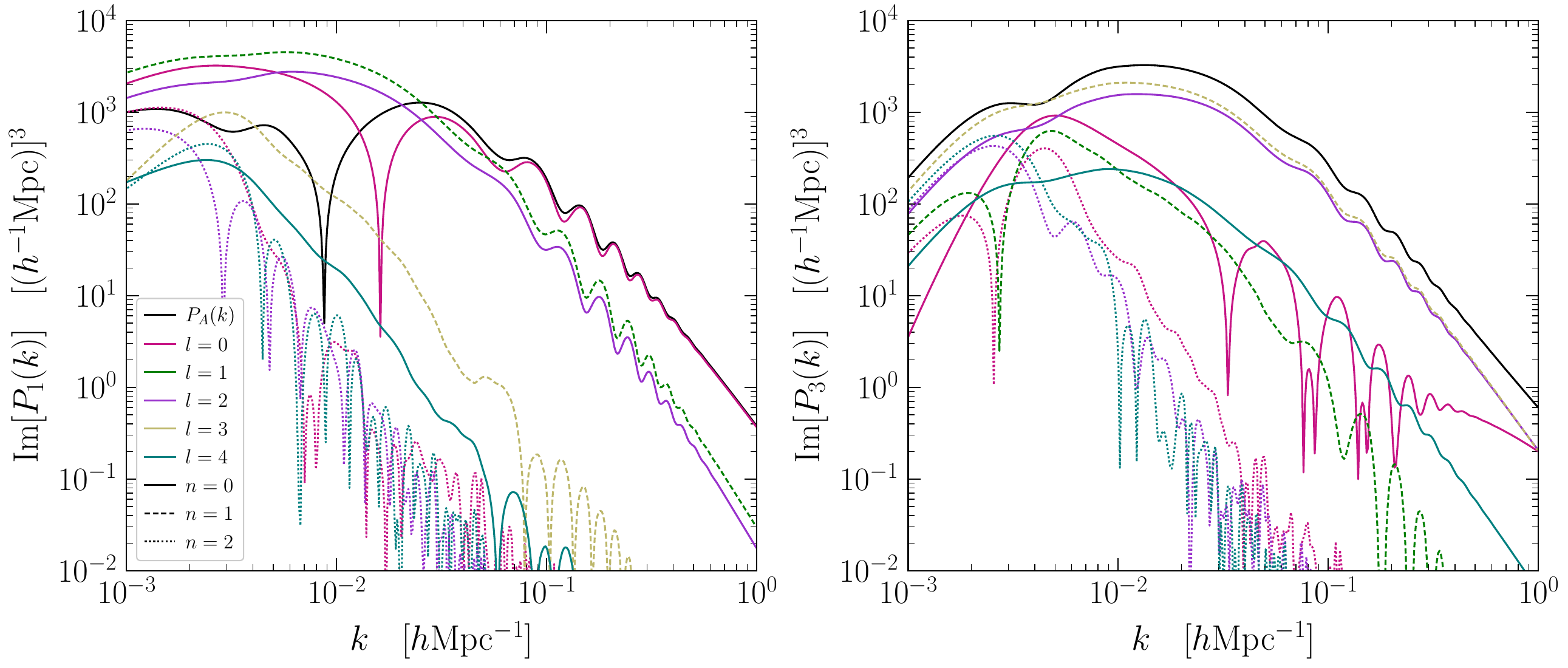"}
\caption{As in Fig.~\ref{fig:SPHEREx_even_multipoles}, but for odd multipoles.
}
\label{fig:SPHEREx_odd_multipoles}
\end{figure*}

We now move on to show results for a low redshift bin $0.2<z<0.5$ in SPHEREx. In Fig.~\ref{fig:SPHEREx_even_multipoles} and~\ref{fig:SPHEREx_odd_multipoles}, we show respectively the even and the imaginary part of the odd multipoles of the convolved power spectra in black solid.  
The colored curves correspond to individual contributions from the various correlation function multipoles $\xi_l^{(n)}$ (see  Eq.~\ref{eq:convpower}), and we restrict ourselves to only $n \le 2$ terms for clarity.

As in Ref.~\cite{Beutler+:2019JCAP...03..040B} for the BOSS survey, the monopole is dominated by the $l=0$, $n=0$ contribution. Indeed for $l=0$ the only non-vanishing terms in Eq.~\ref{eq:convpower} are for $l=L$, and $Q_0^{(0)}(s)$ is the largest of the window multipoles (\reffig{window_multipoles_synthetic_masks_comparison}). The quadrupole at $n = 0$ is the next largest contribution at $20\%$, while $n=1$ contributions are of order $0.5\%$, from the dipole and the octopole. Note that wide-angle contributions can be comparable or larger
than the plane-parallel terms, e.g the $n = 2$ contribution from $l=0$ 
is larger than the $l=4$ contribution at $n = 0$. 
Interestingly, the hexadecapole is highly sensitive to wide-angle effects and receives important contributions from each of the correlation multipoles at all orders $n$. More specifically, at $k \approx 3 \times 10^{-3} \, \iMpch$, we see that the power in the hexadecapole is
spread out among $8$ different terms.


\subsection{Incorporating primordial non-Gaussianity}
\label{sec:nongaussianity}

Primordial non-Gaussianity (PNG) can be incorporated into the wide angle formalism by including a scale-dependent correction to the linear bias, which in turn modifies the correlation function multipoles.
We restrict ourselves here to the modification of the linear galaxy bias.
We consider local PNG such that perturbations of the gravitational potential may be expressed as
\ba
\Phi_{\mathrm{NG}}(\boldsymbol{x})=\varphi(\boldsymbol{x})+f_{\mathrm{NL}}\left(\varphi^{2}(\boldsymbol{x})-\left\langle\varphi^{2}\right\rangle\right) \,,
\ea
where $\varphi$ is an auxiliary primordial Gaussian potential. To first order in $\fnl$, the deviation of the bias at redshift $z$ from the Eulerian halo bias $b_{1,G}(z)$ is \cite{Dalal+:2008PhRvD..77l3514D, Baldauf_2011, Raccanelli:2013dza, Slosar:2008hx}
\ba
b_1(k,z) = b_{1,G}(z) + \fnl\,  b_\phi(z) \, \frac{3\Omega_m H_0^2}{2 c^2 k^2 T(k) D(z)} \,,
\label{eq:bkfnl}
\ea
where $b_\phi(z)=2 \, (b_{1,G}(z) - 1) \, \delta_c$ under the assumption of the universal mass function,
where $\delta_c$ is the critical overdensity for halo collapse (here taken to be the critical density for spherical collapse, $\delta_c = 1.686$), $\Omega_m$ is the present-day  matter density as a fraction of the critical density, $H_0$ the Hubble parameter,
and $T(k)$ is the matter perturbation transfer function (normalized such that $T(k) \to 1$ at small $k$). Here we normalize the growth factor $D$ to unity at $z=0$. 
Consistent with Plank 2018 data which constrains local PNG to have $f_\mathrm{NL} = -0.9 \pm 5.1$ at $68\%$ confidence \cite{planckcollaboration2019planck}, we restrict ourselves to $|f_{\mathrm{NL}}| \le 5$.

When $f_\mathrm{NL}\neq 0$, in the formulae for $\xi_l^{(n)}$ (given in Appendix \ref{app:wide_angle_corrections} along with the plane-parallel results Eqs.~\ref{eq:kaiser_0}--\ref{eq:kaiser_4}) we must replace  $b_1$ and $\beta$ by their scale-dependent expressions, and absorb all scale-dependent prefactors into the Hankel transforms $\Xi_\ell^{(n')}$ (Eq.~\ref{eq:Hankel_transform}).
However, on large scales $b_1(k,z) \sim k^{-2}$ and
$P(k)\sim k$. Therefore, since $j_\ell(ks) \sim k^\ell$ at small $k$, the
integrand of the $\Xi_\ell^{(n')}$ goes as $\sim k^{3+\ell-n}$ for the Gaussian
calculation on large scales, and PNG introduces additional factors, which may grow as fast as $b_1^2 \sim k^{-4}$. 
This means that $\xi_0 ^{(0)}(s)$ in Eq.~\ref{eq:kaiser_0} would diverge for
all $s$.
This divergence does not occur for the other $\xi_l^{(n)}(s)$.

Of course, the correlation function measured in a survey is finite because the mean number density of galaxies inside the survey is measured from the survey itself \cite{Wands_2009}, so that the $k=0$ mode vanishes. This is referred to as the integral constraint or local average effect, which we model following Ref.~\cite{Wands_2009}. We now show that, by moving part of the integral constraint calculation into the $\xi_0 ^{(0)}(s)$, the otherwise divergent integral now converges. 

The convolved power spectrum including the (global) integral constraint is
\ba
\tilde{P}_{l}(k) 
\equiv 
P_{l}(k) 
-
P_0(0)
Q_{l}^{(0)}(k) \,,
\label{eq:icmaintext}
\ea
where we define the (appropriately normalized) Hankel transform of the window multipole $Q_L^{(n)}(s)$ in \cref{eq:Q_window_multipole} as 
\ba
Q_L^{(n)}(k) & \equiv \frac{i^L}{V_\mathrm{survey}}\int_0^{\infty} Q_L^{(n)}(s)j_L(ks)s^2 ds \label{eq:C1maintext}  \,.
\ea
Using Eq.~\ref{eq:power_estimator}, we may rewrite Eq.~\ref{eq:icmaintext} as
\begin{widetext}
\ba
\tilde{{P}}_{l}(k)
&= \bigg[\frac{2l+1}{V_\mathrm{survey}}\int \frac{\mathrm{d} \Omega_k}{ 4\pi} \, \mathrm{d}^3 s_1\,\mathrm{d}^3 s_2\, e^{i \mathbf{k} \cdot \mathbf{s}}\,\left(\langle \delta(\mathbf{s_1})\delta(\mathbf{s_2})\rangle -c_{0}^{0}\right)
W(\mathbf{s_1})W(\mathbf{s_2})\,\mathcal{L}_{l}(\hat{\mathbf{k}}\cdot\hat{\mathbf{s}}_1) \bigg]-c_{L>0}^{n>0}V_\mathrm{survey}Q^{(0)}_{l}(k)\,,
\label{eq:pA}
\ea 
\end{widetext}
where we split the wide-angle expansion of the $k=0$ mode into the contribution $c_{0}^{0}$ from the $L=n=0$ term, and the sum of the remaining convergent terms $c_{L>0}^{n>0}$:
\ba
{P}_{0}(0)
&= \sum_{
L,n}\begin{pmatrix}
L & L & 0 \\
0 & 0 & 0
\end{pmatrix}^2\int \mathrm{d}s\, s^{n+2} \,\xi_{L}^{(n)}(s) Q_L^{(n)}(s) \nonumber \\&=V_\mathrm{survey} \left [c_{0}^{0}+c_{L>0}^{n>0}\right].
\label{eq:p00maintext}
\ea

Now
it suffices to calculate the bracketed term 
in Eq.~\ref{eq:pA} using the wide-angle expansion Eq.~\ref{eq:convpower} with a modified correlation function monopole $\tilde{\xi}^{(0)}_{0}(s) \equiv \xi^0_{0}(s)-c_{0}^{0}$. 
Combining Eq.~\ref{eq:kaiser_0} and Eq.~\ref{eq:Hankel_transform} we find
\ba
\tilde{\xi}_0^{0}(s)= &\int\!{\frac{k^2\,\mathrm{d}k}{2\pi^2}}\left( 1 + \frac{2}{3}  \beta +  \frac{1}{5} \beta^2 \right)b_1^2P(k) \nonumber \\ 
&\times \bigg[ j_0(ks)-Q_0^{(0)}(k)\bigg ]\,.   
\label{eq:ksi_bias}
\ea
If the region where the window function is non-vanishing has characteristic linear size $r_*$, then the term in the brackets is quadratic in $k$ on scales  $k \ll r_*^{-1}$, so that the integrand goes as $k$, and Eq.~\ref{eq:ksi_bias} converges.

\begin{figure*}
\includegraphics[width=1\textwidth]{"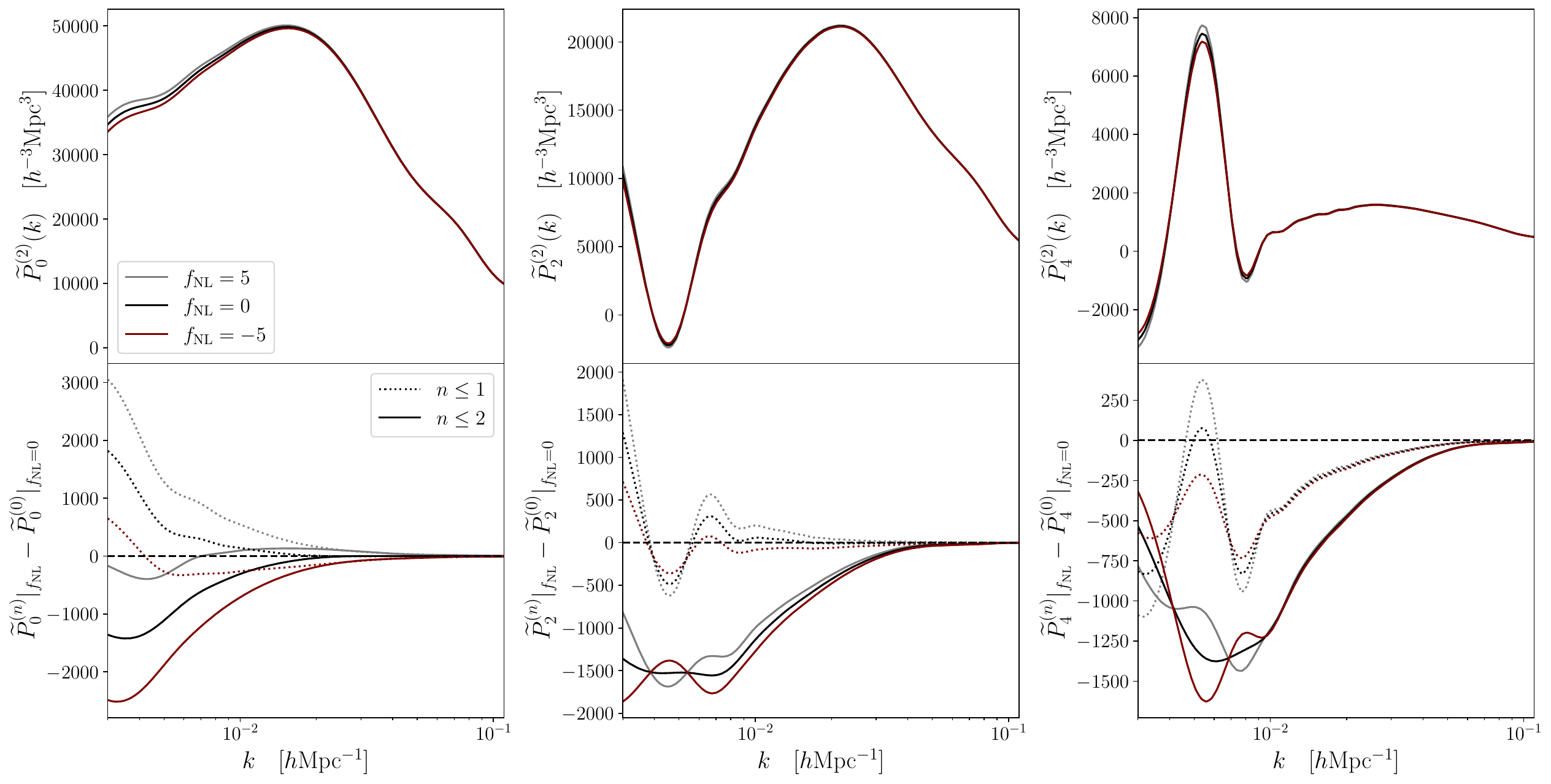"}
\caption{Convolved power spectrum with wide-angle effects and PNG for the SPHEREx redshift bin $0.2<z<0.4$. Here $b_{1,G}=1.8$. Upper panel: convolved power spectrum even multipoles computed to order $n=2$ for various values of $f_{\mathrm{NL}}$. Lower panel: deviation of the convolved power multipoles with their value for $n=0 \,, f_{\mathrm{NL}}=0$. 
}
\label{fig:SPHEREx_endpoint_fnl_deviation_ordernfrom0fnl0_comparison}
\end{figure*}

\begin{figure*}
\includegraphics[width=0.8\textwidth]{"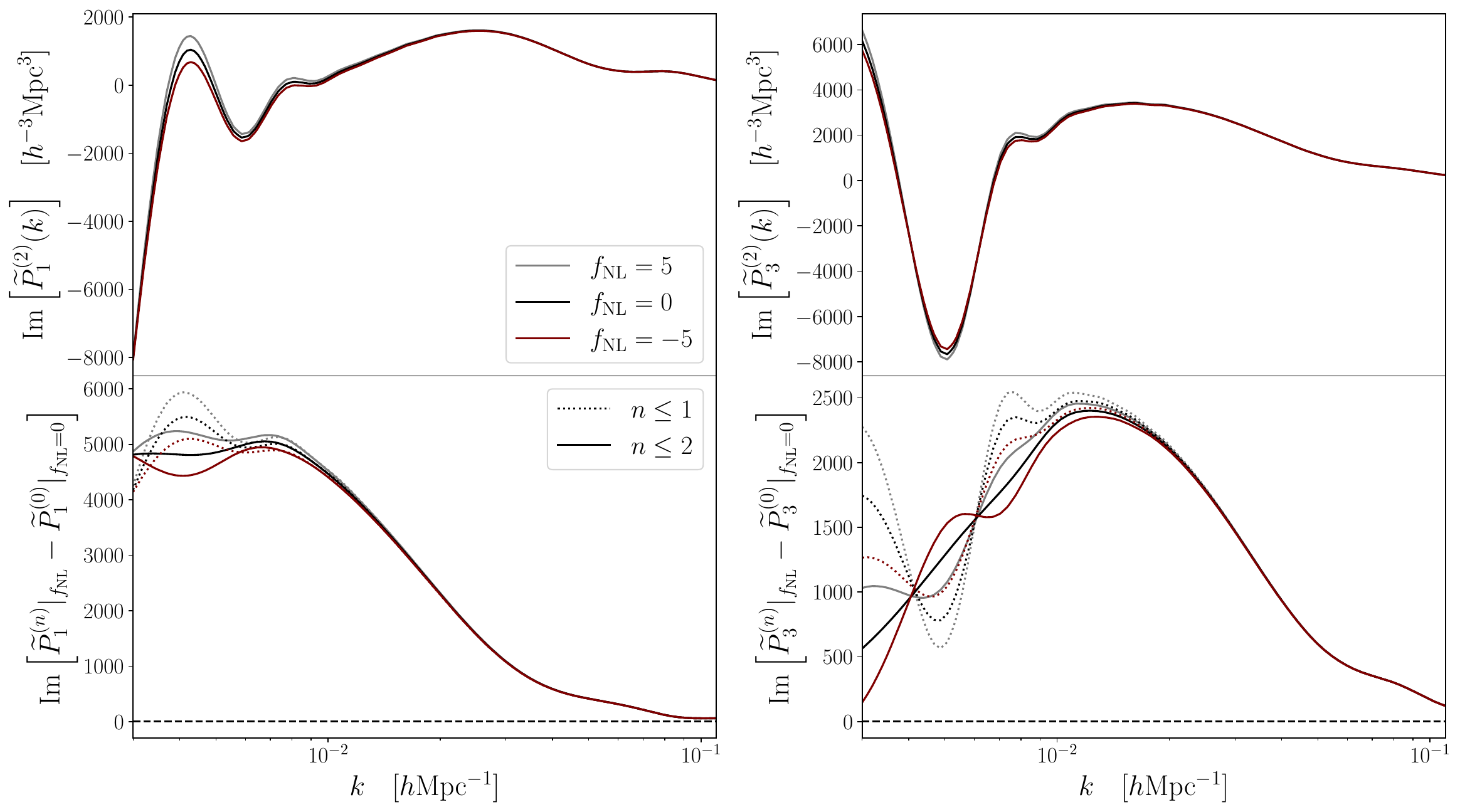"}
\caption{As in Fig.~\ref{fig:SPHEREx_endpoint_fnl_deviation_ordernfrom0fnl0_comparison}, but for the dipole and octopole.
}
\label{fig:SPHEREx_endpoint_fnl_deviation_ordernfrom0fnl0_comparison_oddmultipoles}
\end{figure*}
We now show the even and odd power spectrum multipoles accounting for the integral-constraint for nonzero values of $f_{\mathrm{NL}}$ in Figs.~\ref{fig:SPHEREx_endpoint_fnl_deviation_ordernfrom0fnl0_comparison} and~\ref{fig:SPHEREx_endpoint_fnl_deviation_ordernfrom0fnl0_comparison_oddmultipoles} respectively. 
In the top panel, we show the signal including wide-angle effects up to $n\leq2$ for $\fnl = 0, \pm 5$. 
In the lower panel we plot the fractional difference of these curves up to order $n=1,2$ with respect to the curve for $n=0$, $f_{\mathrm{NL}}=0$. 
As the scale-dependent bias increases (decreases) with scale according to the sign of $f_{\mathrm{NL}}$, the power is increased (suppressed) relative to the non-Gaussian case for $f_{\mathrm{NL}}>0$ ($f_{\mathrm{NL}}<0$).
The effect of $f_{\mathrm{NL}}=5$ nearly cancels wide-angle effects at order $n=2$ for the monopole (gray solid curve in the left bottom panel).

Compared to the monopole, all higher multipoles are more strongly affected by wide-angle effects than by primordial local non-Gaussianity. This is because $\ell>0$ are primarily sourced by unrelated physics like redshift space distortions and window convolution. In principle, this should allow breaking this WA-$\fnl$ degeneracy. However, we leave a more detailed analysis to a future paper.

\section{Simulations}
\label{sec:simulations}
In this section we use simulated galaxy catalogs to show that the perturbative framework is insufficient to accurately model the power spectrum multipoles to the percent level down to scales $ k \sim 10^{-3} \, \iMpch$ for the SPHEREx redshift bin $0.2 < z < 0.5$ (we will elaborate on this choice of redshift range in Section. \ref{sec:comparison}).
On smaller scales we establish which order $n$ is sufficient.  We begin by describing our setup for generating mock galaxy catalogs and applying the Yamamoto estimator on the mocks to measure the power spectrum multipoles.
We then discuss the comparison between the simulation measurements and the perturbative modeling framework.
From here on we fix $f_\mathrm{NL}=0$. 

\subsection{Log-normal galaxy catalogs}
\label{sec:log-normals}
We simulate the galaxy density field that would be observed by SPHEREx using mock log-normal galaxy catalogs \citep{1991MNRAS.248....1C}, largely following the procedure described in Ref.~\cite{2017_Agrawal}. We use log-normal simulations 
since they are inexpensive and give mock catalogs with a known power spectrum on large scales. We do not model the Fingers-of-God effect in our log-normal mocks (nor in the wide-angle framework of the previous sections), as it only becomes significant at scales smaller than where wide-angle effects are expected to arise \cite{2018JCAP...02..039L}. 

To generate the catalogs, we draw a Gaussian field $G(\mathbf{k})$ with power spectrum
\begin{equation}
    P_G(\mathbf{k})
    = \int \dd^3r e^{-i\mathbf{k}\cdot\mathbf{r}} \xi_G(\mathbf{r}) \,,
\end{equation}
where $\xi_G(\bm{r})$ is the log-space correlation function
\ba
    \xi_G(\mathbf{r}) \equiv \ln(1 + \xi(\mathbf{r}))\,,
\ea
with $\xi(\mathbf{r}) =\int d^3 k e^{i \mathbf{k} \cdot\mathbf{x}} P(k)$ the configuration space correlation function. The log-transformed density field is then
\begin{equation}
    G(\mathbf{x}) \equiv \ln(1 + \delta(\mathbf{x}))- \langle \ln(1 + \delta(\mathbf{x}))\rangle_\mathbf{x} \,,
    \label{eq:lognorm}
\end{equation}
where we invoke the ergodic theorem to write the ensemble average as a spatial average, indicated by the $\vx$ suffix. After Fourier-transforming $G(\mathbf{k})$ to configuration space, the log-transformed overdensity field $\delta(\mathbf{x})$ is given by
\begin{equation}
    \delta(\mathbf{x}) = e^{-\frac{1}{2}\sigma_G^2 + G(\mathbf{x})} -1 \,,
    \label{eq:delta_ln}
\end{equation}
where $\sigma_G^2 = \langle G(\mathbf{x})^2 \rangle_\mathbf{x}$ ensures that the average overdensity of the simulated catalog vanishes: $\langle \delta(\mathbf{x}) \rangle_\mathbf{x} = 0$.

In each cell of a cubic grid  of size $N_{\mathrm{log}}^3$, we Poisson sample
galaxy number counts from the density contrast (Eq.~\ref{eq:delta_ln}). To
incorporate RSD, the galaxy velocity field is generated according to the linear theory
result Eq.~\ref{eq:vk}, and the redshift space positions are computed
with Eq.~\ref{eq:doppler_shift} \cite{2017_Agrawal}. We then impose periodic
boundary conditions such that galaxies remain inside the simulation volume.

Typically, the grids of the simulation and the estimator are aligned to
match each other. In real space that means that all galaxies of one cell in the
simulation can be put in one cell in the estimator, so that the voxel window
effect cancels \cite{2017_Agrawal}. If the grids are misaligned, however, then
we observe a significant voxel window effect that suppresses power on
intermediate to small scales.

Since RSD distort the regular grid into an irregular grid that is imprinted in
the distribution of the simulated galaxies, it is generally impossible to align
the estimator grid with the imprinted grid. However, in order to reduce the
sensitivity of the measured power spectrum to such grid misalignment by
smoothing the density field, we employ the following procedure (which to our knowledge is not used elsewhere in the literature):
First, we uniformly draw a position inside the grid cell, then we again
uniformly draw from a box the same size as a grid cell, but centered around the
new position. 

As a result, the probability distribution of the grid cell is a
pyramid that extends half to all its neighboring grid cells, so that when
accounting for the neighboring cells also having the shape of a pyramid, the
full probability distribution is a trilinear interpolation of the density
field. Similarly, we choose the velocity of the galaxy as the average velocity
field over a small region around the final position.

This smoothing procedure suppresses power on small scales in the same manner as
the voxel window effect in the estimator (see \cite{Jing:2005ApJ...620..559J,
Sefusatti+:2016MNRAS.460.3624S} for a discussion on the voxel window effect in
estimators). Crucial for a successful correction is that our procedure leads to
a mock catalog that is not very sensitive to the grid misalignment. Since the
reduction in power on small scales due to this smoothing is similar to that due
to the \emph{cloud-in-cell} grid assignment scheme when estimating the
fluctuation field (both are convolutions of two 3D top-hats -- one top-hat for the shape of the galaxy, one for the shape of the voxel), we may term this the  \emph{reverse cloud-in-cell} (RCIC)
galaxy-drawing scheme. Indeed, to correct we merely need to increase the power
in the voxel window modeling of the estimator.

Lastly, to model the survey window function, galaxies lying outside of the
redshift bin of interest or angular mask are removed.

\subsection{Yamamoto estimator}
\label{sec:Yamamoto_estimator}
Power spectrum estimation beyond the plane-parallel approximation requires an estimator that uses a variable LOS.
The power spectrum estimator we use is a version of the Yamamoto estimator \citep{Yamamoto+:2006PASJ...58...93Y,
Bianchi:2015oia, Scoccimarro:2015bla}. The novelty of our implementation is in the correction of power leakage between multipoles, $\mu$-leakage, for which we generalize the result in Ref.~\cite{2017_Agrawal} to a varying LOS
\footnote{Our estimator code is written in Julia (\url{https://julialang.org}) and will be made publicly accessible following a subsequent dedicated publication.}. We leave details of our estimator implementation to \cref{app:yamamoto}, and give a brief overview here.

We define the galaxy fluctuation field by \citep{Feldman:1993ky, Hand17_FFT}
\ba
\label{eq:Fr}
F(\mathbf{r})
&=
w_{\textrm{FKP}}(\mathbf{r})
\, \frac{n_g(\mathbf{r}) - \alpha' n_s(\mathbf{r})}{\sqrt{A}}
\,,
\ea
where $n_g(\mathbf{r})$ and $n_s(\mathbf{r})$ are the number density for the
galaxy catalog and a synthetic  catalog of uniformly sampled galaxies,
respectively. The ratio of the number of galaxies to the number of objects in
the synthetic catalog is denoted by $\alpha'$.
In this work we have $n_s(\vr)$ sampled from a constant uniform distribution inside the survey volume. Hence, we can choose $w_\mathrm{FKP}(\vr)=1$.
The normalization $A$ is chosen such that the expectation for $\hat P_l(k)$
coincides with $P_l(k)$ for modes well within the survey volume
\cite{Feldman:1993ky}. It is given by
\ba
\label{eq:A}
A
&=
\int\frac{\dd^3r}{V_\mathrm{survey}}
\,w^2_\mathrm{FKP}(\vr)\,\bar n^2(\vr)
=
\bar n^2\,,
\ea
where the last equality follows for our simulations, since we use a constant
average number density $\bar n$.
Rather than measuring $\bar n$ from the simulations, we use the input to the simulations, which allows us to avoid the local average effect.

Substituting the observed fluctuation field \cref{eq:Fr}, the estimator \cref{eq:power_estimator} becomes \cite{Hand17_FFT}
\ba
    \hat{P}_l(k)
    &=
    \frac{1}{V_\mathrm{survey}}\int \frac{\dd\Omega_k}{4\pi} \, F_{l}(\mathbf{k}) F_0^*(\textbf{k})
    - P_l^{\rm{noise}}(k)\,,
\label{eq:yamamoto}
\ea
where 
\ba
    \label{eq:Fl}
    F_{l}(\mathbf{k})
    &=
    (2l+1) \int \dd^3r_2 \, e^{-i \mathbf{k}   \cdot \mathbf{r}_2} 
    \, \mathcal{L}_l(\hat{\mathbf{k}}\cdot\hat{\mathbf{r}}_2)
    \, F(\mathbf{r}_2)
    \\
    &=
    4\pi \sum_{m} Y_{l m}(\hat{\mathbf{k}}) \int \dd^3r_2 \, e^{-i \mathbf{k} \cdot \mathbf{r}_2} 
    \, Y^*_{l m}(\hat{\mathbf{r}}_2) \, F(\mathbf{r}_2) \,,
    \label{eq:Fl_Yamamoto}
\ea
where $\Omega_k$ is the solid angle in Fourier space. \cref{eq:Fl_Yamamoto} can
be evaluated using the fast-Fourier-transform algorithm \citep{Bianchi:2015oia,
Scoccimarro:2015bla, Hand17_FFT}\footnote{Code for the spherical harmonics was
generated using the computer algebra package
\texttt{SymPy}\footnote{\url{https://www.sympy.org}}.}. In \cref{eq:yamamoto} we
also subtracted the shot noise
\ba
    \label{eq:PlNoise}
    P_l^{\rm{noise}}(k)
    \equiv
    \delta^K_{l 0}
    \,\frac{(1+\alpha')}{A\,V_\mathrm{survey}}
    \int  d^3 r \, w_{\mathrm{FKP}}^2(\vr)  \, \bar{n}(\vr)
    \,.
\ea

We assign the objects in the log-normal galaxy catalog to a cubic mesh grid of size $N_\mathrm{est}^3$ using the nearest-grid-point (NGP) method. We also test with the cloud-in-cell (CIC) interpolation method, 
but do not find an improvement. Crucial is that the lognormals are created with our RCIC method, as described in the previous section. Interpolating the galaxies onto a regular grid of voxels introduces a systematic in the multipole estimation. This is the voxel window effect that we correct for similar to Refs.~\cite{Jing:2005ApJ...620..559J, Sefusatti+:2016MNRAS.460.3624S, 2017_Agrawal}.

In order to avoid aliasing from periodic boundary conditions in
\cref{eq:Fl_Yamamoto}, we use a zero-padded simulation box that is 50\% larger
in linear extent than the minimum box required to fit the specified survey
volume. Furthermore, as the angular integrals Eq.~\ref{eq:Fl_Yamamoto} are performed as a sum over discrete angular bins, discretization induces a leakage of power between $\ell$ modes ($\mu$-leakage). 
To account for this effect, we opt for the approach first noted in Ref.~\cite{2017_Agrawal}. However, we generalize their approach specifically for a windowed varying LOS estimator. The full derivation is presented in \cref{sec:muleakage_varying_LOS}. While the correction is important in the plane-parallel limit, we find that it is negligible for a spherically symmetric window, see \cref{sec:muleakage_fixed_LOS,sec:mu_discussion} for a more extensive discussion of this effect.

\subsection{Wide-angle effects in simulations}
\label{sec:sims_full_sky}
\begin{figure}
    \centering
    \includegraphics[width=0.5\textwidth]{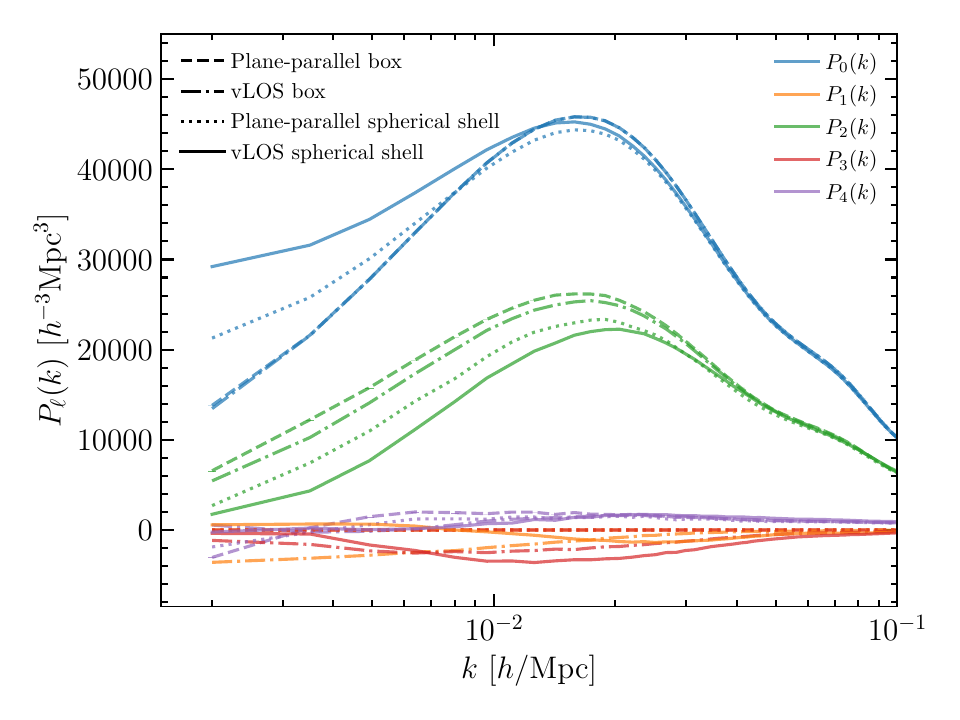}
    \caption{
    Mean $P_\ell(k)$ from $10^4$ simulations (imaginary part shown for odd $\ell$).
    The line style indicates the observer position and survey geometry: \emph{Plane-parallel} refers to placing the observer far from the survey for the LOS calculation only, and \emph{vLOS} indicates the observer in the center of the simulation box; \emph{box} refers to a full simulation box with periodic boundary conditions that imply replication across the universe, and \emph{spherical shell} refers to a survey geometry that is a thick spherically symmetric shell around the center of the box. The colors indicate the multipole.
    The monopole is insensitive to WA effects for a full box with periodic boundaries, as the dashed blue and dash-dotted blue lines overlap. Once a window is introduced, the WA effect is clearly visible as an enhancement of the power on large scales, as shown by the dotted blue and solid blue lines.
    The quadrupole is different in all four cases, and the other multipoles are shown for completeness only.
    See the main text for caveats concerning the $\ell>0$ multipoles.
    }
    \label{fig:sims_ep_pp_win_nowin}
\end{figure}

\begin{figure*}
  \centering
  \includegraphics[width=0.49\textwidth]{"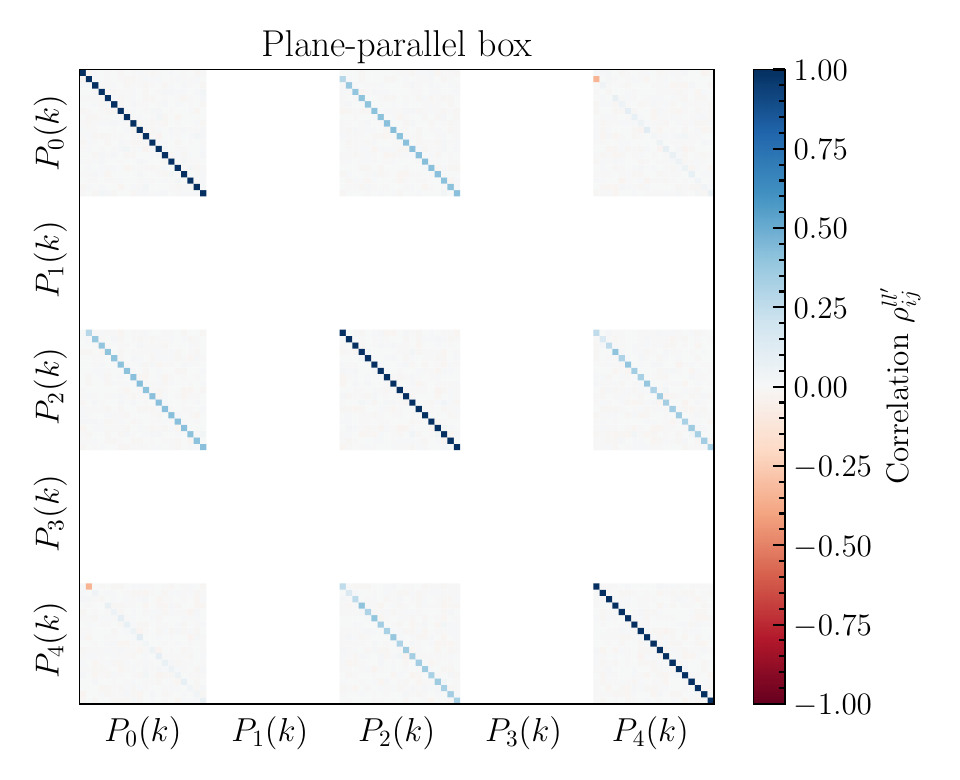"}
  \includegraphics[width=0.49\textwidth]{"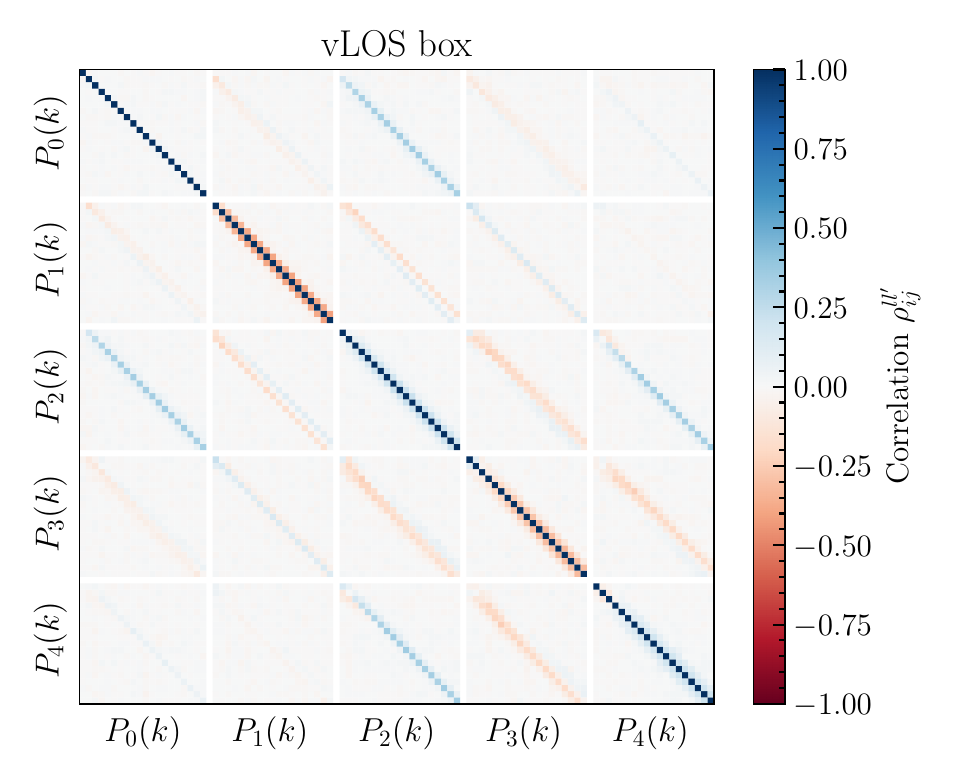"}
  \includegraphics[width=0.49\textwidth]{"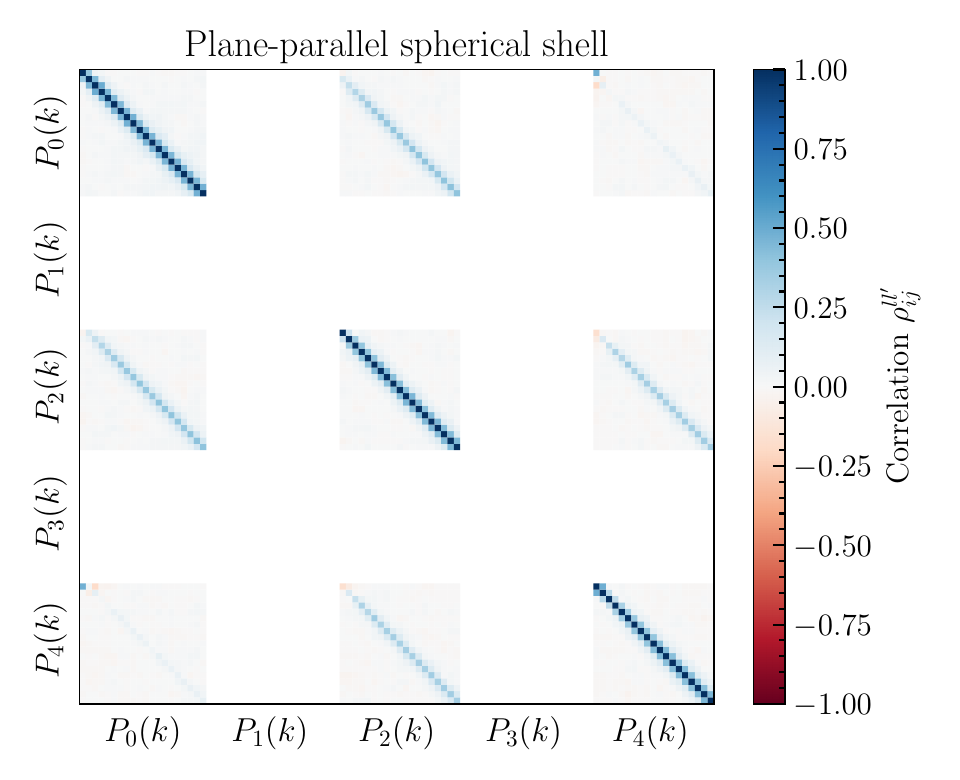"}
  \includegraphics[width=0.49\textwidth]{"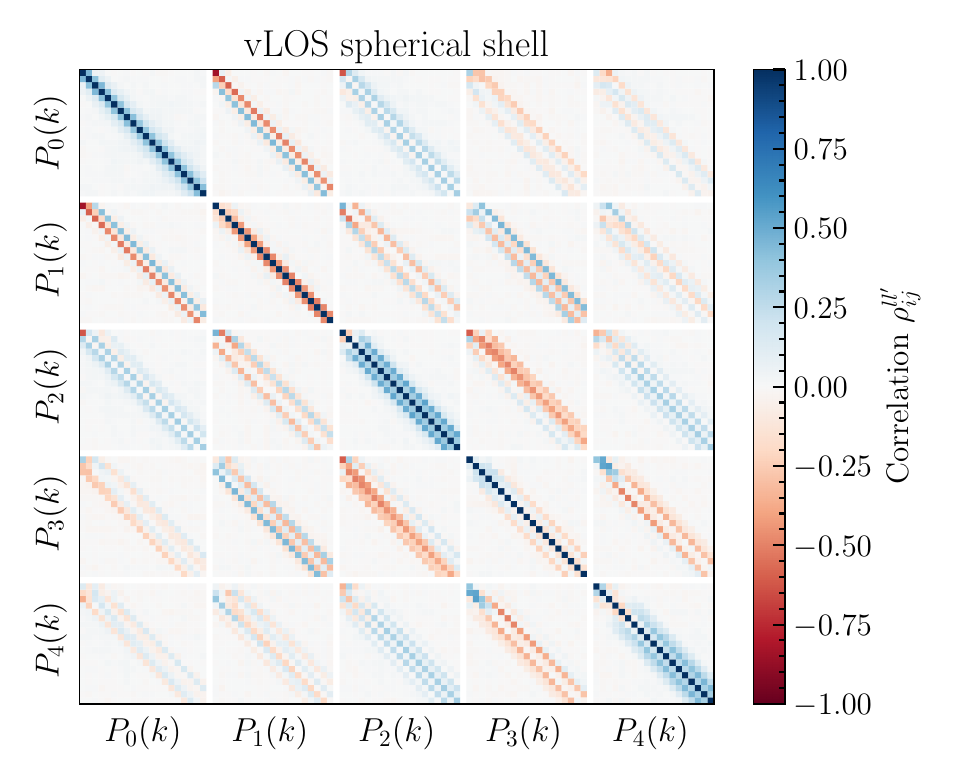"}
  \caption{
  The power spectrum correlation matrices are shown for the four cases of \cref{fig:sims_ep_pp_win_nowin} measured from $10^4$ galaxy catalogs.
  Top left: Plane-parallel box without window and periodic boundary conditions. Since the odd multipoles vanish, they are masked-out (see \cref{foot:odd_multipoles_pp}).
  Top right: Same survey geometry as in the upper left, but with a varying LOS.
  Bottom left: Spherical shell $0.2<z<0.5$ survey geometry with fixed LOS for the plane-parallel approximation.
  Bottom right: Spherical shell $0.2<z<0.5$ survey geometry with observer in the center (full-sky), corresponding to \cref{fig:SPHEREx_even_multipoles_model_sim_comparison_0.2_0.5,fig:SPHEREx_odd_multipoles_model_sim_comparison_0.2_0.5}.
  In each panel, only the 20 largest modes are shown, ranging from $k=0$ to $k\simeq 0.03\,\iMpch$. The $k=0$ mode is meaningless for $\ell>0$ and is therefore masked.
  }
  \label{fig:SPHEREx_correlation_matrix_0.2_0.5}
\end{figure*}

For comparison between modeling and simulations we choose a survey geometry that has both a large volume and relatively large wide-angle effects. Wide-angle effects are most important for a full-sky survey, which restricts simulations to lower redshifts due to computational constraints. On the other hand, the variance in the measured power spectrum is lower the larger the volume. Thus, as a middle ground between these two trade-offs, we choose our fiducial survey as a full-sky survey with redshift $0.2\leq z\leq 0.5$.
We use a simulation box with mesh size $N^3=512^3$, cubic box side length $L=4000 \, \Mpch$, and number density $\bar n=3\times 10^{-4}\,h^3\,\mathrm{Mpc}^{-3}$. For simplicity, the survey geometry is applied without any other selection function effects. We use $10^4$ simulations and apply the estimator to each. 
The regime of validity of our simulations can be inferred from the right panel of \cref{fig:pkl_pp_nowin} in the appendix, and is further discussed in \cref{sec:mu_discussion}. In particular, the hexadecapole must be interpreted with caution, as it shows significant deviation from the expectation. 
However, both monopole and quadrupole show percent-level agreement, and it is those we focus on in this section.

Before we delve into a comparison between the perturbative calculation and the simulations, in this section let us explore the wide-angle effect with the simulations alone. \cref{fig:sims_ep_pp_win_nowin} shows the mean of the $10^4$ simulations in four different variations constructed from two different treatments of the LOS (plane-parallel or varying LOS) and two different survey geometries (box or spherical shell), explained in the following. In each variation, we start with the same lognormal realization.

The LOS enters as follows. Before the survey geometry is applied, we use the first-order velocity from the lognormal simulation to shift the galaxies according to \cref{eq:doppler_shift} and thus add RSD to the simulation. In order to simulate the plane-parallel LOS, we artificially place the observer far outside the analysis box ($10^8\,\Mpch$) so that the LOS is approximately constant. However, this is done for the purpose of calculating the LOS only, and this LOS is used both for applying the RSD and in the Yamamoto estimator. However, it does not change the actual galaxy distribution which thus has the same intrinsic clustering in real space. In the varying LOS (vLOS) case we place the observer in the center of the box.

By \emph{box} geometry we mean that the full Fourier box used in the analysis is filled with galaxies from the lognormal simulation, and periodic boundary conditions are applied. The \emph{spherical shell} geometry means that galaxies are only within a thick spherical shell at the center of the box, with the inner and outer radii corresponding to redshifts $z=0.2$ and $z=0.5$ for an observer at the center.

Because for the \emph{box} geometry we fill the entire analysis box and apply periodic boundary conditions, that means that we assume replication of the box throughout the entire universe. In the case of varying LOS, this implies that the direction of the RSD is always towards the center of each replicated box, and we caution the reader to be aware of this when interpreting \cref{fig:sims_ep_pp_win_nowin}. On the other hand, when the spherical shell survey geometry is applied, our Fourier space box is large enough that it is effectively zero-padded and we can ignore the periodic boundaries.

\cref{fig:sims_ep_pp_win_nowin} shows that the wide-angle effect in the monopole is relevant only when the survey geometry is applied.
For the higher multipoles the wide-angle effect shows in the simulations regardless. In particular, the odd multipoles vanish identically for both plane-parallel cases, whereas they are nonzero with the varying LOS.

The hexadecapole should be interpreted with caution in this plot, as it differs significantly from the expectations in the plane-parallel case (e.g., see right panel of \cref{fig:pkl_pp_nowin}). However, despite this caveat, we find significantly better agreement in the vLOS case with our modeling, as will become apparent later in \cref{sec:comparison}.

The multipole correlation matrices estimated from the simulations are shown in Fig.~\ref{fig:SPHEREx_correlation_matrix_0.2_0.5}, for the same four cases as in \cref{fig:sims_ep_pp_win_nowin}. The correlation coefficient is defined by
\ba
\rho^{ll'}_{ij} = \frac{C^{ll'}_{ij}}{\sqrt{C^{ll'}_{ii}C^{ll'}_{jj}}}\,,
\ea
where $C^{ll'}_{ij}\equiv C^{ll'}(k_i,k_j)$ is the covariance matrix, defined via
\ba
C^{ll'}(k_m,k_n)\equiv& \,\langle\Delta P_l(k_m) \,
\Delta P_{l'}(k_n)\rangle,
\ea
where $\Delta P_l(k)\equiv \, \widehat{P}_l(k) - P_l(k) $ for even $l$, and the imaginary component of the preceding expression for odd $l$.
Since the plane-parallel simulations have vanishing odd multipoles, they are masked-out in the correlation matrix\footnote{\label{foot:odd_multipoles_pp}Due to our procedure of emulating the plane-parallel case by moving the simulation box far from the observer, the odd multipoles are not quite identically zero, but suppressed by a factor $10^{-7}$ compared to the monopole. Therefore, the raw correlation matrix estimated from the simulations contains strong cross-correlations between the even and odd multipoles even in the plane-parallel case.} in Fig.~\ref{fig:SPHEREx_correlation_matrix_0.2_0.5}. Similarly, the $k=0$ mode for the $\ell>0$ multipoles is masked out. 

Comparing the plane-parallel box with the vLOS box in \cref{fig:SPHEREx_correlation_matrix_0.2_0.5} shows that wide-angle effects introduce small off-diagonal terms for the even multipoles. It also shows that the correlations with odd multipoles generally have many off-diagonal terms. It is well-known that the window introduces off-diagonal correlations between neighboring $k$-modes. In the wide-angle case, however, these off-diagonal terms are significantly enhanced, as shown in the bottom right plot of \cref{fig:SPHEREx_correlation_matrix_0.2_0.5}.

Similar to \citet{Beutler+:2019JCAP...03..040B} we find the odd multipoles strongly correlated with the even multipoles, which is expected from the fact that the $\xi_l^{(n)}$ for odd $l$ are entirely given by combinations of even multipoles (see \cref{app:wide_angle_corrections} for details).
We find this to be true both with and without a window, though the correlations are stronger with a window. This highlights the importance of the covariance modeling that will be needed for future surveys.

\section{Comparison between perturbative calculation and simulations}
\label{sec:comparison}

\begin{figure*}
\includegraphics[width=1\textwidth]{"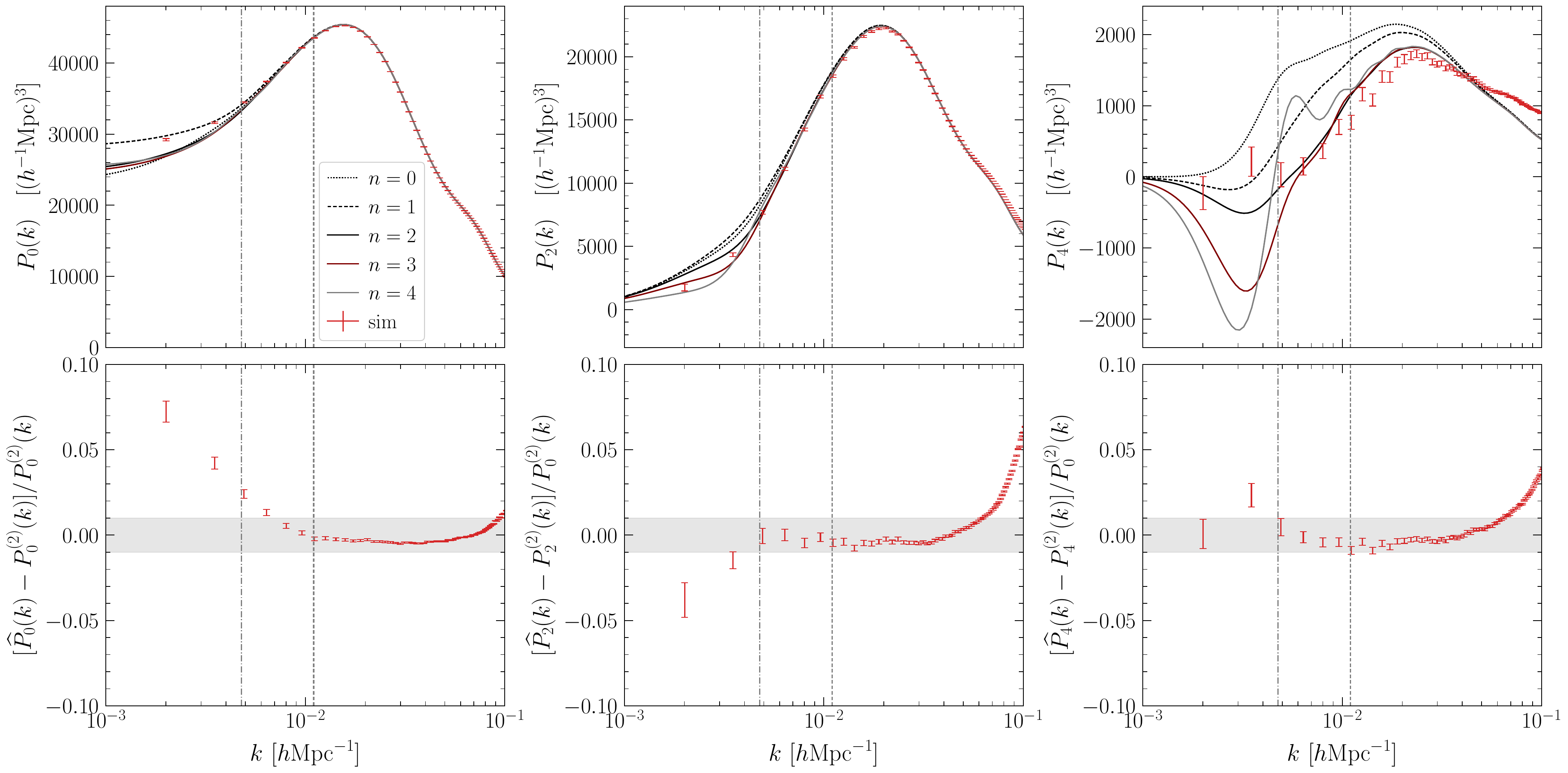"}
\caption{ 
Yamamoto estimator measurement of the even power spectrum multipoles for a spherically symmetric mask and redshift bin $0.2<z<0.5$ with flat galaxy bias $b_1=1.455$ (red error bars),
averaged over $10^4$ galaxy catalogs with error bars indicating the standard deviation on the mean. We compare to the perturbative result at orders $n=0,1,2,3,4$ (black curves).
The lower panels show the relative deviation between the catalog measurement and the perturbative result at order $n=2$, normalized to the perturbative result for the monopole. Deviations of $\pm 1\%$ are indicated by the gray bands.
In each panel the vertical gray dotted (dash-dotted) line corresponds to $k=k_{c_1}\equiv 2\pi/\chi_\mathrm{min}$ ($k=k_{c_2}\equiv2\pi/\chi_\mathrm{max}$), defined in \cref{eq:kc1_kc2}.
Since this is the scale at which some (all) galaxy pairs have $x_s>1$
this is an estimate for the scale at which the perturbative approach becomes invalid for any $n$.
On small scales the figure shows voxel window effects that are not fully captured by the standard voxel window correction. 
\\
}
\label{fig:SPHEREx_even_multipoles_model_sim_comparison_0.2_0.5}
\end{figure*}

\begin{figure*}
\includegraphics[width=0.8\textwidth]{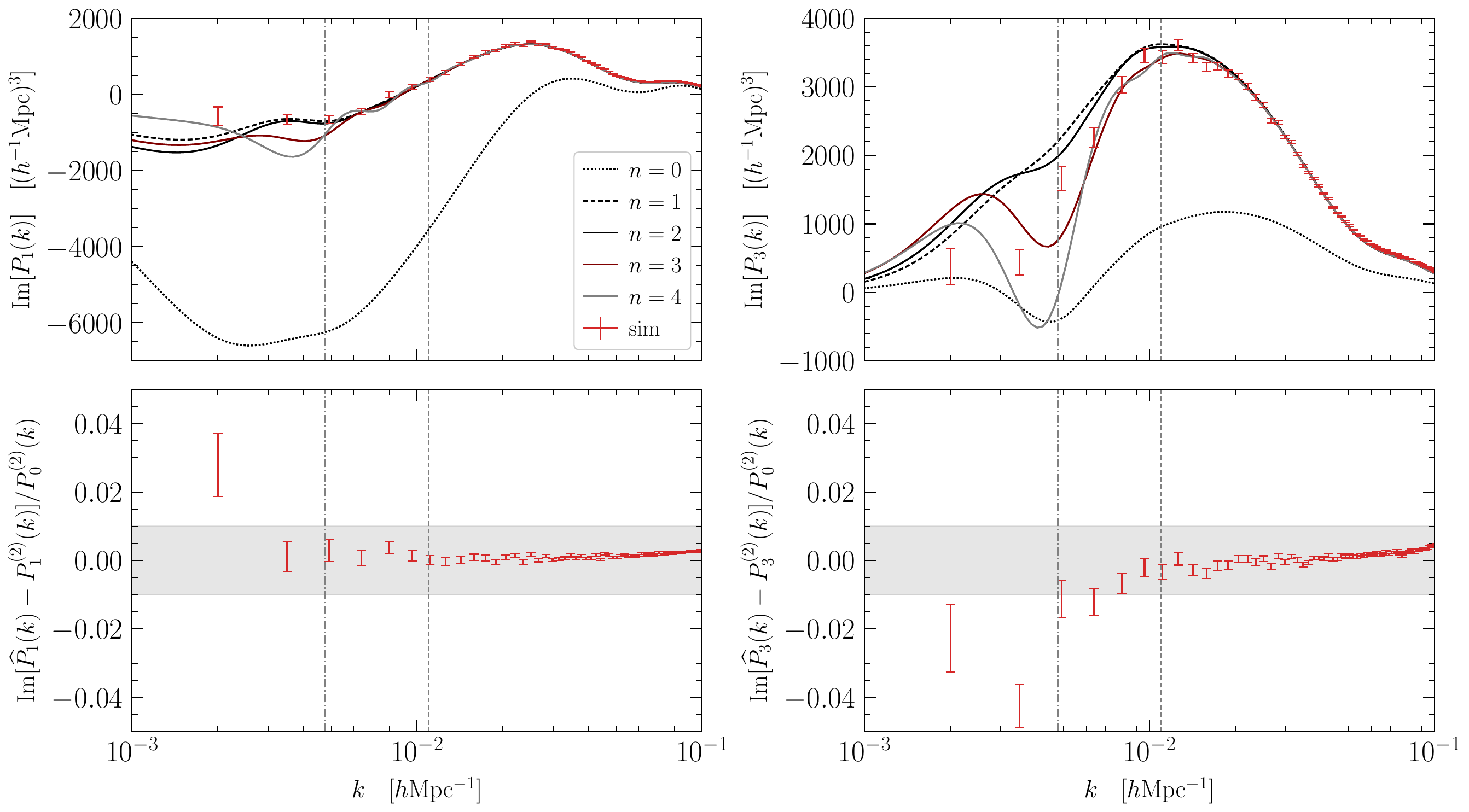}
\caption{As in Fig.~\ref{fig:SPHEREx_even_multipoles_model_sim_comparison_0.2_0.5}, but for the dipole and octopole.
}
\label{fig:SPHEREx_odd_multipoles_model_sim_comparison_0.2_0.5}
\end{figure*}

In this section we compare our perturbative results from \cref{sec:results} with the simulations from \cref{sec:simulations}. We also extend to fourth-order perturbative terms, and we estimate the regime of validity for the perturbative approach, verified by simulations.

In \cref{fig:SPHEREx_even_multipoles_model_sim_comparison_0.2_0.5,fig:SPHEREx_odd_multipoles_model_sim_comparison_0.2_0.5} the simulations are compared with the perturbative results up to fourth-order. 
The curves show the perturbative calculations (include the $\alpha$-term in \cref{eq:alpha}), and the red dots show the simulations averages with $1\sigma$ standard deviations. The standard deviation on the mean is estimated from the $10^4$ simulations.
The monopole shown in the left panels of \cref{fig:SPHEREx_even_multipoles_model_sim_comparison_0.2_0.5} shows an excess in the simulations. Similarly, the quadrupole shows a suppression on the very largest modes~\footnote{The small-scale discrepancy is likely due to the incomplete correction of the voxel window and it is not focus of this paper.}. Although the hexadecapole is unreliable in the plane-parallel approximation, the figure shows good agreement in the vLOS case. In the lower panels, we show the residuals between the simulations and the $n \le 2$ perturbative result. Here we normalize the residuals by the monopole, as it is the leading term in the multipole expansion and it is also the leading contribution to the variance.

Fig.~\ref{fig:SPHEREx_odd_multipoles_model_sim_comparison_0.2_0.5} shows sub-percent level agreement of the dipole and octopole power spectra over a wide range of scales when using the perturbative terms up to $n=2$.

\cref{fig:SPHEREx_even_multipoles_model_sim_comparison_0.2_0.5,fig:SPHEREx_odd_multipoles_model_sim_comparison_0.2_0.5} also show the perturbative approach up to fourth order ($n=4$). For the monopole, we do not find a sufficiently large change over second order to account for the discrepancy. For the multipoles $\ell=1,2,3,4$ we find the $n=4$ large-scale correction to be of similar amplitude or larger than the $n=2$ correction, which suggests that the perturbative expansion is breaking down.

Furthermore, in the residual plots of \cref{fig:SPHEREx_even_multipoles_model_sim_comparison_0.2_0.5,fig:SPHEREx_odd_multipoles_model_sim_comparison_0.2_0.5}, we plot two vertical lines at wave numbers
\ba
\label{eq:kc1_kc2}
k_{c_1} &= \frac{2\pi}{\chi_\mathrm{min}}
&\text{and}&&
k_{c_2} &= \frac{2\pi}{\chi_\mathrm{max}}\,,
\ea
where $\chi_\mathrm{min}$ and $\chi_\mathrm{max}$ are the inner and outer radii of the redshift bin, respectively.
For separations  
$s>2\pi/k_{c_1}$, there exist galaxy pairs with separation $s$ larger than than the LOS distance $d$, thus for which the expansion parameter $x_s=s/d$ exceeds unity and the wide-angle expansion is not valid.
For $s>2\pi/k_{c_2}$, all galaxy pairs with separation $s$ have $x_s>1$.
Hence, $k_{c_2}$ is an estimate of the threshold below which the perturbative result should not be valid at any $n$. See Appendix  \ref{sec:counting_galaxy_pairs} for more details.

\section{Conclusion}
\label{sec:conclusion}

In this work we explore the applicability of a simple perturbative approach to modeling wide-angle effects in the power spectrum multipoles to next-generation galaxy surveys, with a particular focus on SPHEREx. 
Our main result is that the perturbative expansion becomes invalid when the separation $s$ between a pair of galaxies exceeds the LOS distance $d$ to the pair,
and in \cref{fig:SPHEREx_even_multipoles_model_sim_comparison_0.2_0.5,fig:SPHEREx_odd_multipoles_model_sim_comparison_0.2_0.5} we show this explicitly by comparing to simulations.
We expect the perturbative approach to break down when a large fraction of the galaxy pairs have $s >d$
and this leads to the heuristic given in \cref{eq:kc1_kc2} for a full-sky survey. 

Furthermore, we elaborate on the wide-angle effect for several realistic masks in \cref{fig:survey_comparison_end-point_vs_kaiser}, and for different redshift bins for SPHEREx in \cref{fig:SPHEREx_effect_of_redshift}. We also extend the formalism (which works well on intermediate scales) to include local primordial non-Gaussianity (PNG). To avoid divergences in the intermediate steps of the calculation, the local average effect should be accounted for. We show the results in \cref{fig:SPHEREx_endpoint_fnl_deviation_ordernfrom0fnl0_comparison,fig:SPHEREx_endpoint_fnl_deviation_ordernfrom0fnl0_comparison_oddmultipoles}. We find that the PNG is somewhat degenerate with wide-angle effects for the monopole. However, all $\ell>0$ multipoles show different behaviors for wide-angle and $\fnl$ effects.

After generating a suite of $10^4$ SPHEREx-like lognormal catalogs, we detect the wide-angle effect purely from simulations in \cref{fig:sims_ep_pp_win_nowin}, where we show that the monopole is affected only in the presence of a window. The corresponding correlation matrices are shown in \cref{fig:SPHEREx_correlation_matrix_0.2_0.5}, and the correlation matrices acquire additional off-diagonal terms in the wide-angle case compared to the plane-parallel case.

Our results motivate the use of non-perturbative methods to model the multipoles. Until recently, such methods have been less explored due to their computational cost, as they typically involve higher dimensional integrals with highly oscillatory integrands \citep{Castorina:2019hyr,Castorina+:2022JCAP...01..061C}.  Fortunately, some of the computational hurdles are overcome in the companion paper Ref.~\cite{Wen+:2024}, permitting a fast, exact calculation of the multipoles showing sub-percent-level agreement with simulations for the monopole. The method used in Ref.~\cite{Wen+:2024} is to compute correlation functions of the density contrast using a basis which diagonalizes the Laplacian operator in spherical coordinates (i.e, to compute the $2$-point function of the spherical Fourier-Bessel (SFB) modes \cite{{Gebhardt:2021mds,Benabou:2023ldb, Gebhardt:2021tme, Gebhardt:2023kfu,Castorina:2019hyr}} of the galaxy distribution), and then apply a transformation to obtain the multipoles. 

Our work leaves a number of interesting directions to be explored. For example, it would be illuminating to repeat the analysis in this paper for the galaxy bispectrum. Indeed, the measurement of the bispectrum by surveys such as SPHEREx will be crucial for constraining PNG, and is complementary to probes of the scale-dependent bias from the power spectrum. Ref.~\cite{Pardede:2023ddq} expresses the bispectrum via a series expansion in the angular separation between galaxies, analogous to the one used in this paper. They find that wide-angle effects in the monopole can mimic local PNG with $\fnl = \mathcal{O}(0.1)$. Further work is needed to cross-examine such perturbative frameworks with simulations, or directly compare them to non-perturbative methods such as the  bispectrum in tomographic spherical harmonics space \cite{Dio_2016} or the SFB bispectrum \cite{Benabou:2023ldb, bertacca}. 

Ultimately, a Fisher forecast of $\fnl$ constraints for SPHEREx should be performed, including information from both the power spectrum and bispectrum, in a framework that fully accounts for wide-angle effects. 
Such an analysis can then be compared to existing ones which work in the plane-parallel approximation such as that in Ref.~\cite{Heinrich:2023qaa}. Note that, once wide-angle corrections are fully accounted for, other effects relevant on large scales may still bias the estimation of $\fnl$. These include general relativistic (GR) effects which are integrated along the line-of-sight, such as lensing magnification,  Integrated Sachs-Wolfe, and Shapiro time-delay \citep{Bernardeau+:2002PhR...367....1B,Jeong+:2015CQGra..32d4001J}. While some progress has been made toward quantifying the biases on PNG measurements introduced by these effects \cite{Castorina:2021xzs}, accounting for them brings computational difficulties, and a fully self-consistent quantification of degeneracies between PNG and GR effects is yet to be completed. 

On the simulations side,  dedicated simulations of the galaxy distribution observed by SPHEREx will play a central role in validating such studies. To go beyond the simplistic simulated catalogs we use for power spectrum estimation in this work, a number of ingredients should be incorporated, including a realistic SPHEREx survey mask and radial selection function, and photometric redshift errors (see Ref.~\cite{Feder:2023rqg} for recent encouraging work on SPHEREx catalogs with realistic photo-$z$ uncertainties). 

Simulations will also be essential in informing theory priors that enter in the modeling of PNG on large scales. For example, in studying the effect of the scale-dependent bias on the power spectrum multipoles, we rely on the assumption of the universality of the halo mass function, which fixes the relation between the Eulerian halo bias $b_{1,G}$ and the PNG parameter $b_\phi$ (see Eq.~\ref{eq:bkfnl}). While this assumption is widely adopted \citep{Dore:2014cca, dePutter:2014lna,Raccanelli:2015oma}, there is in fact a large uncertainty on $b_\phi$, and an inaccurate assumption on the value of this parameter can significantly bias power spectrum constraints on $\fnl$ \cite{Barreira:2022sey}. Dedicated galaxy formation simulations such as those in Refs.~\citep{Barreira:2020kvh, Hadzhiyska:2024kmt} will help reduce this uncertainty.

In the present era of precision cosmology, these and other ongoing efforts to tackle the challenges in cosmological modeling and data analysis presented by the next-generation sky surveys are of utmost importance. With careful work, SPHEREx measurements of the galaxy distribution over a volume much larger than previously accessible will provide us with a unique means to distinguish between models of inflation.

\acknowledgements
We thank Robin Wen, Florian Beutler, Emanuele Castorina,  Andrew Robinson, and Gabor Racz for useful discussions, and Dida Markovič for providing Roman and Euclid survey footprints.
This work used resources of the Texas Advanced Computing Center at The University of Texas at Austin.
J.B is supported in part by the DOE Early Career Grant DESC0019225.
For part of this work, H.G was supported by an appointment to the NASA Postdoctoral Program at the Jet Propulsion Laboratory.
Part of this work was done at the Jet Propulsion Laboratory, California Institute of Technology, under a contract with the National Aeronautics and Space Administration (80NM0018D0004).
\bibliography{psmultipole.bib}

\appendix
\onecolumngrid

\clearpage
\section{Counting galaxy pairs outside of the perturbative regime}
\label{sec:counting_galaxy_pairs}

\begin{figure*}[ht]
\includegraphics[width=0.5\textwidth]{"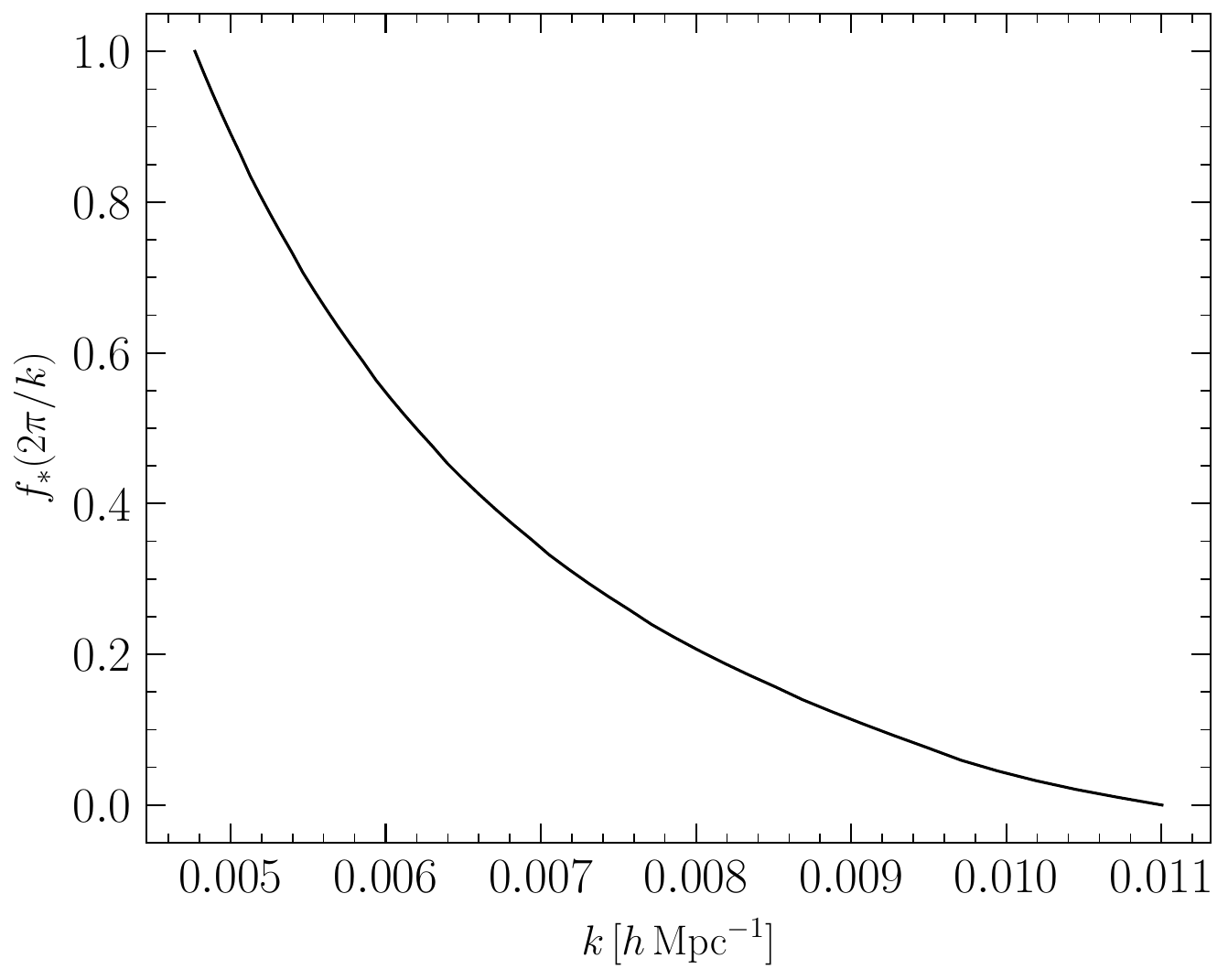"}
\caption{For a spherically symmetric survey window with redshift range $[z_\mathrm{min},z_\mathrm{max}]=[0.2,0.5]$, the proportion of galaxy pairs separated by a distance approximately equal to $s$, for which $x_s>1$.}
\label{fig:f60}
\end{figure*}

Here we seek to estimate the fraction of galaxy pairs lying within the survey window  separated by a distance approximately equal to $s$, which also satisfy $x_s>1$, such that the wide-angle expansion is no longer valid. 

We assume a spherically symmetric window and a redshift bin $[\chi_\mathrm{min},\chi_\mathrm{max}]$ (as in Eq.~\ref{eq:window_function}). To simplify, let us assume the galaxy number density is constant in space. 
Let $V(S,D)$ denote the 6-dimensional volume in the space of galaxy pairs with coordinates $(\mathbf{s}_1,\mathbf{s_2})$ with $|\mathbf{s}_1-\mathbf{s}_2|<S$ and $D>s_1$. Then the proportion of pairs with separation in the interval $[s,s+ds]$ which also have $x_s>1$ is $f_*(s) \equiv (\partial V(S,s)/\partial S) / (\partial V(S, \chi_\mathrm{max})/\partial S ) |_{S=s}$. Further,
\ba
V(S,D) &= \int d\mathbf{s}_1W(\mathbf{s}_1) \int d\mathbf{s}_2 
W(\mathbf{s}_2) \Theta(S-|\mathbf{s}_1-\mathbf{s}_2|) \Theta(D-|\mathbf{{s}_1}|)\nonumber \\
&=(4\pi)(2\pi)  \int_{\chi_\mathrm{min}}^{\chi_\mathrm{max}} s_2^2ds_2 \int_{\chi_\mathrm{min}}^{D} s_1^2ds_1  \int_{-1}^{1} dx \, \Theta\left(x- \frac{s_1^2+s_2^2-S^2}{2s_1s_2}\right) \,,
\ea
where $\Theta$ is the Heaviside step function, and to obtain the second line we perform the angular integral over $\hat{\mathbf{s}}_1$ and rewrite $\mathbf{s_1}\cdot\mathbf{s_2}$ using the law of cosines. Differentiating inside the integrand, the step function becomes a Dirac delta, giving
\ba
\frac{\partial V(S,D)}{\partial S} \Big|_{S=s} 
&=8\pi^2s\int_{\chi_\mathrm{min}}^{\chi_\mathrm{max}} s_2ds_2 \int_{\chi_\mathrm{min}}^{D} s_1ds_1\mathbf{1}_{[-1,1]}\left(\frac{s_1^2+s_2^2-s^2}{2s_1s_2}\right) \,,
\ea
which may be integrated numerically. We plot $f_{*}(s)$ in terms of the Fourier conjugate $k=2\pi/s$ in Fig.~\ref{fig:f60} for $[z_\mathrm{min},z_\mathrm{max}]=[0.2,0.5]$. The $x$-axis limits correspond to $s=\chi_\mathrm{max}$ ($s=\chi_\mathrm{min}$), for which all (none) of the galaxy pairs with separation $s$ have $x_s>1$. Note that in Fig.~\ref{fig:SPHEREx_even_multipoles_model_sim_comparison_0.2_0.5} (which assumes the same redshift bin), the monopole deviates from the $n=2$ expansion at the percent-level for $k\lesssim 6 \times 10^{-3}\, \iMpch$, when  $f_{*}(s) \sim 50\%$.

\section{Wide-angle corrections to correlation function multipoles}
\label{app:wide_angle_corrections}

Here we list the modifications to the correlation function multipoles up to order four in the wide-angle expansion. To our knowledge the $n=3,4$ terms are not written explicitly elsewhere in the literature. 
\subsection{Corrections up to second order}

In the bisector LOS, symmetry about the LOS requires that $n=1$ corrections to the correlation function multipoles vanish. The $n=2$ corrections are expressed in terms of the $\Xi_l^{(0)}(s)$ (Eq.~\ref{eq:Hankel_transform}) as \cite{Reimberg_2016}
\ba
  \xi_0^{{\mathrm{bis.}},(2)}(s) &= -\frac{4\beta^2}{45}b_1^2\Xi_0^{(0)}(s)-\frac{\beta(9+\beta)}{45}\ b_1^2\Xi_2^{(0)}(s)  \label{eq:correlation_multipoles_bisector_order0_1} \,,\\
  \xi_2^{{\mathrm{bis.}},(2)}(s) &=  \frac{4\beta^2}{45}\ b_1^2\Xi_0^{(0)}(s) + \frac{\beta(189+53\beta)}{441}\ b_1^2\Xi_2^{(0)}(s) -\frac{4\beta^2}{245}\ b_1^2\Xi_4^{(0)}(s)  \label{eq:correlation_multipoles_bisector_order0_2} \,,\\
  \xi_4^{{\mathrm{bis.}},(2)}(s) &= -\frac{8\beta(7+3\beta)}{245}\ b_1^2\Xi_2^{(0)}(s) + \frac{4\beta^2}{245}\ b_1^2\Xi_4^{(0)}(s) \,.
\label{eq:correlation_multipoles_bisector_order0_3}
\intertext{The correlation function multipoles in the end-point LOS at orders $n=1,2$ are then given in terms of the above and the $n=0$ results Eqs.~\ref{eq:kaiser_0}--\ref{eq:kaiser_4} by \cite{Beutler+:2019JCAP...03..040B}}
\label{eq:order1term0}
   \xi_1^{{\rm ep},(1)}(s) &=-\frac{3}{5}\xi_2^{(0)}(s) \,, \\
   \xi_3^{{\rm ep},(1)}(s) &= \frac{3}{5}\xi_2^{(0)}(s)-\frac{10}{9}\xi_4^{(0)}(s)\,, \\
    \xi_5^{\mathrm{ep},(1)}(r)&=\frac{16}{63} \beta^2 \Xi_4^0(s) \,, \\
  \xi_0^{{\rm ep},(2)}(s) &= \xi_0^{{\mathrm{bis.}},(2)}(s) + \frac{1}{5}\xi_2^{(0)}(s) \,,\\
  \xi_2^{{\rm ep},(2)}(s) &=  \xi_2^{{\mathrm{bis.}},(2)}(s) -  \frac{2}{7}\xi_2^{(0)}(s)  + \frac{5}{7}\xi_4^{(0)}(s) \,,\\
  \xi_4^{{\rm ep},(2)}(s) &=  \xi_4^{{\mathrm{bis.}},(2)}(s)+ \frac{3}{35}\xi_2^{(0)}(s) - \frac{90}{77}\xi_4^{(0)}(s) \,, \\
  \xi_6^{{\rm ep},(2)}(s) &= \frac{64 \beta^2}{231}\Xi_4^{(0)}(s) 
  \,.
\label{eq:correlation_multipoles_end-point}
\intertext{
Taking into account the galaxy selection function, the correlation function multipoles up to $\mathcal{O}\left(x_s^2\right)$ assuming $\alpha$ is independent of redshift receive corrections which are  suppressed by powers of the distance to the galaxy pair $d$ ~\cite{Beutler+:2019JCAP...03..040B}. They are
}
\xi^{\mathrm{ep},(2)}_0(s) & \ni b_1^2 s^{-2}\frac{4 \beta^2}{3}\Xi_0^{(2)}(s)+\frac{2}{3 }s^{-1} b_1^2 \beta(1-\beta) \Xi_1^{(1)}(s) \,, \\
\xi^{\mathrm{ep},(2)}_2(s) & \ni-\frac{8}{3} b_1^2 s^{-2} \beta^2 \Xi_2^{(2)}(s)-\frac{8}{15 } b_1^2 s^{-1} \beta(5+\beta)  \Xi_1^{(1)}(s)+\frac{4}{5 } b_1^2 \beta^2 s^{-1}\Xi_3^{(1)}(s) \,.
\ea
Note that in Eq.~\ref{eq:convpower} the products $s^{n+2}\xi^{(n)}_{l}(s)$ now have a small $s$ dependence like $s^2 \Xi^{(n')}_{l}(s) \sim s$, so no divergences are introduced by the selection function terms.
\subsection{Third and fourth order corrections}
It is straightforward, using Eq.~4.7 and Eq.~4.11 of Ref.~\cite{Reimberg_2016} to write down the higher order terms, including those due to the galaxy selection function. We give them below at orders $n=3,4$. For brevity we  suppress factors of $b_1^2$ which multiply each term in the $\xi^{(n)}_l(s)$.
\ba
\xi_1^{{\rm ep},(3)}(s) &= \left(\frac{2 \beta}{5}-\frac{26 \beta^2}{75}\right) \Xi^{(0)}_0(s) +\frac{4 \beta^2}{3  s^2}\Xi^{(2)}_0(s) + \left(\frac{6 \beta}{35}-\frac{226 \beta^2}{735}\right) \Xi^{(0)}_2(s) +\frac{4 \beta^2}{3  s^2}\Xi^{(2)}_2(s)  +\left(-\frac{136 \beta^2}{3675}\right)\Xi^{(0)}_4(s) \,,\\
\xi_3^{{\rm ep},(3)}(s) &= \left(-\frac{16 \beta}{15}-\frac{8 \beta^2}{25}\right) \Xi^{(0)}_0(s) + \left(\frac{8 \beta}{45}+\frac{8 \beta^2}{105}\right) \Xi^{(0)}_2(s) +\left(\frac{328 \beta^2}{5775}\right)\Xi^{(0)}_4(s) \,,\\
\xi_5^{{\rm ep},(3)}(s) &=  \left(-\frac{64 \beta}{63}-\frac{64 \beta^2}{147}\right)\Xi^{(0)}_2(s) + \left(-\frac{608 \beta^2}{1911}\right)\Xi^{(0)}_4(s) \,,\\
\xi_7^{{\rm ep},(3)}(s) &= \frac{128 \beta^2}{429}\Xi^{(0)}_4(s) \,, \\
\xi_0^{{\rm ep},(4)}(s) &= \left(\frac{2 \beta}{45}+\frac{38 \beta^2}{225}\right) \Xi^{(0)}_0(s) -\frac{4 \beta^2}{9  s^2}\Xi^{(2)}_0(s) + \left(\frac{8 \beta}{315}+\frac{304 \beta^2}{2205}\right) \Xi^{(0)}_2(s) -\frac{4 \beta^2}{9  s^2}\Xi^{(2)}_2(s) +\left(\frac{148 \beta^2}{11025}\right)\Xi^{(0)}_4(s) \,, \\
\xi_2^{{\rm ep},(4)}(s) &= \left(\frac{32 \beta}{63}-\frac{148 \beta^2}{315}\right) \Xi^{(0)}_0(s) +\frac{16 \beta^2}{9  s^2}\Xi^{(2)}_0(s) + \left(\frac{2 \beta}{9}-\frac{26 \beta^2}{63}\right) \Xi^{(0)}_2(s) +\frac{16 \beta^2}{9  s^2}\Xi^{(2)}_2(s) +\left(-\frac{172 \beta^2}{3465}\right)\Xi^{(0)}_4(s) \,,  \\
\xi_4^{{\rm ep},(4)}(s) &= \left(-\frac{128 \beta}{105}-\frac{64 \beta^2}{175}\right) \Xi^{(0)}_0(s) + \left(\frac{32 \beta}{165}+\frac{32 \beta^2}{385}\right) \Xi^{(0)}_2(s) +\left(\frac{1592 \beta^2}{25025}\right)\Xi^{(0)}_4(s) \,, \\
\xi_6^{{\rm ep},(4)}(s) &=   \left(-\frac{256 \beta}{231}-\frac{256 \beta^2}{539}\right)\Xi^{(0)}_2(s) + \left(-\frac{8384 \beta^2}{24255}\right)\Xi^{(0)}_4(s) \,, \\
\xi_8^{{\rm ep},(4)}(s) &= \frac{2048 \beta^2}{6435}\Xi^{(0)}_4(s)
\,.
\label{eq:n_3_4_ksi}
\ea
For ease of reproducibility we provide a \texttt{Mathematica} notebook 
\href{https://github.com/joshua-benabou/wide_angle_expansion
}
{\faGithub} 
to compute $\xi^{(n)}_l(s)$ symbolically for arbitrary $(l,n)$.

\section{Yamamoto estimator implementation details}
\label{app:yamamoto}
In this Appendix we detail how we correct for $\mu$-leakage in our implementation of the Yamamoto estimator. As it turns out, $\mu$-leakage is important in our implementation of the estimator whenever the window function deviates from spherical symmetry about the observer.

We give a brief explanation of the $\mu$-leakage effect in
\cref{sec:muleakage_fixed_LOS}, a detailed derivation in
\cref{sec:muleakage_varying_LOS}, and we discuss in
\cref{sec:mu_discussion} with an example in \cref{fig:pkl_pp_nowin}.

\subsection{Useful formulae}

Here we list formulae that we employ in the subsequent subsections. 

Legendre polynomials decompose into spherical harmonics via
\ba
\label{eq:legendre_ylmylm}
\mathcal{L}_\ell(\khat\cdot\rhat)
&=
\frac{4\pi}{2\ell+1}
\sum_m Y^*_{\ell m}(\khat) \, Y_{\ell m}(\rhat)\,.
\ea
Products of spherical harmonics with the same argument admit the basis expansion \citep{Varshalovich+:1988qtam.book.....V}
\ba
\label{eq:YlmYlm}
Y_{lm}(\qhat)\,Y_{l'm'}(\qhat)
&=
\sum_{LM}
(-1)^{M}
\sqrt{\frac{(2l+1)(2l'+1)(2L+1)}{4\pi}}
\begin{pmatrix}
  l & l' & L \\
  0 & 0  & 0
\end{pmatrix}
\begin{pmatrix}
  l & l' & L \\
  m & m' & -M
\end{pmatrix}
\,Y_{LM}(\qhat)\,,
\ea
where $\tj{l}{l'}{l''}{m}{m'}{m''}$ is a Wigner-3j symbol, which obeys the orthogonality relation \citep{NIST:DLMF}
\ba
\label{eq:3j_orthogonality}
(2L_1+1)\sum_{mm'}
\begin{pmatrix}
  l & l' & L_1 \\
  m & m' & M_1
\end{pmatrix}
\begin{pmatrix}
  l & l' & L_2 \\
  m & m' & M_2
\end{pmatrix}
&=
\delta^K_{L_1 L_2}
\delta^K_{M_1 M_2}\,.
\ea

\subsection{\texorpdfstring{$\mu$}{Mu}-leakage in the plane-parallel limit}
\label{sec:muleakage_fixed_LOS}

Before we derive a more general result for the $\mu$-leakage matrix of the endpoint estimator, we first briefly review it in the context of a fixed LOS, as explored in Ref.~\cite{2017_Agrawal} (their Appendix D).

The galaxy power spectrum multipoles $P_\ell(k)$ are defined as the coefficients in the Legendre polynomial expansion,
\begin{equation}
    P(k,\mu) = \sum_{\ell = 0}^{\infty} P_{\ell}(k)\mathcal{L}_{\ell}(\mu)\,,
    \label{eq:D1}
\end{equation}
where $\mu=\khat\cdot\rhat$ is the cosine of the angle between the Fourier vector $\vk$ and the LOS $\rhat$.
The power spectrum multipoles can then be estimated via
\ba
    \label{eq:plk_integral}
    \hat P_l(k)
    &=
    \sum_\ell\left[
    (2\ell+1)\int_{-1}^1\frac{\dd\mu}{2}\,\mathcal{L}_l(\mu) \,\mathcal{L}_\ell(\mu)\right]
    P_\ell(k)
    \,.
\ea
Integrating exactly, the term in brackets is the unit matrix $\mathcal{M}_{l\ell}(k)=\delta^K_{l\ell}$. 
In practice, the estimator first estimates $P(k_x,k_y,k_z)=P(k,\mu,\phi)$ and
then estimates the average within a Fourier-shell of thickness $\Delta k$
around $k$,
\ba
    \hat P_l(k)
    &=
    \frac{2l+1}{4\pi k^2 \Delta k}
    \int_{k-\frac{\Delta k}{2}}^{k+\frac{\Delta k}{2}}\dd k
    \int_{-1}^1\dd\mu
    \int_0^{2\pi}\dd\phi
    \,\mathcal{L}_l(\mu)
    \,P(k,\mu,\phi)\,.
\ea
This is evaluated as a discrete sum
\ba
    \hat P_l(k)
    &=
    \frac{2l+1}{V_\mathrm{shell}(k)}\sums{j \, s.t. \\ |\vk_j|\sim k}
    k_F^3\,\mathcal{L}_l(\mu_j)
    \,P(k_j,\mu_j,\phi_j)\,,
    \label{eq:Plk_estimate}
\ea
where $V_\mathrm{shell}(k)\simeq 4\pi k_F^2 \Delta k$ is the Fourier-space
volume of the shell, and $k_F$ is the fundamental mode, and the sum is over
all the modes within the shell.
Therefore, the term in brackets in Eq.~\ref{eq:plk_integral} is
\ba
    \label{eq:Mlm_pp}
    \mathcal{M}_{l\ell}(k)
    &=
    \frac{2\ell+1}{N_\mathrm{modes}(k)}
    \sums{j \, s.t. \\ |\vk_j|\sim k}
    \mathcal{L}_l(\mu_j) \,\mathcal{L}_\ell(\mu_j)\,,
\ea
where $N_\mathrm{modes}(k)=V_\mathrm{shell}(k)/k_F^3$ is the number of Fourier-cells used in the sum. The $k$-dependence is due to the varying number and distribution of cells across $k$-shells. The matrix $\mathcal{M}_{l\ell}(k)$ encodes the angular (``$\mu$'') leakage of one $\ell$-mode to another due to the discretization error. Therefore, the initial estimate $\hat P_l(k)$ can be improved by calculating this $\mu$-leakage matrix and inverting it for each $k$ bin. That is, Eq.~\ref{eq:plk_integral} is inverted and becomes
\ba
    P_\ell(k)
    &=
    \sum_l \mathcal{M}^{-1}_{\ell l}(k)\,\hat P_l(k)\,,
\ea
which gives a more accurate estimate of the power spectrum multipoles than Eq.~\ref{eq:Plk_estimate}.

\subsection{\texorpdfstring{$\mu$}{Mu}-leakage for a varying LOS with a window}
\label{sec:muleakage_varying_LOS}
Let us now generalize the preceding calculation to a varying LOS and allow for a nontrivial window function.
The correlation function is
\ba
\langle F(\vr_1) F(\vr_2) \rangle
&= W(\vr_1) W(\vr_2)\,\xi(\vr_1 - \vr_2)
= W(\vr_1) W(\vr_2)\int \frac{\dd^3 k}{(2\pi)^3} e^{i \vk \cdot(\vr_1 - \vr_2)} \sum_{\ell}\mathcal{L}_{\ell}(\khat \cdot \rhat_1)P_{\ell}(k)\,,
\label{eq:corr_pkl}
\ea
where we expressed the correlation function $\xi$ in terms of the multipoles of its Fourier transform.
In the following we will solve Eq.~\ref{eq:corr_pkl}  for $P_\ell(k)$ while keeping track of discretization error. 
To do so, we drop the ensemble average on the left-hand side and we apply the operations that the Yamamoto estimator performs on $F(\vr_1)F(\vr_2)$ (see \cref{eq:yamamoto,eq:Fl}), i.e multiply by a Legendre polynomial $(2L+1)\mathcal{L}_L(\qhat\cdot\rhat_1)$, and Fourier transform over $\vr_1$ and $\vr_2$. After rearranging terms, we get
\ba
&\frac{2L + 1}{V_\mathrm{survey}}
\int \dd^3r_1 \, e^{-i\vq \cdot \vr_1} \, \mathcal{L}_L(\qhat \cdot \rhat_1) \, F(\vr_1)
\int \dd^3r_2 \, e^{i\vq \cdot \vr_2} \, F(\vr_2)
\vs
&=
\sum_{\ell}
\int \frac{\dd^3 k}{(2\pi)^3} \,P_{\ell}(k)
\left[\frac{2L+1}{V_\mathrm{survey}}\int \dd^3r_1 \, e^{-i(\vq-\vk) \cdot \vr_1}
\,W(\vr_1)
\,\mathcal{L}_{L}(\qhat \cdot \rhat_1)
\,\mathcal{L}_{\ell}(\khat \cdot \rhat_1)
\right]\left[
\int \dd^3r_2 \, e^{i(\vq-\vk) \cdot \vr_2}
\,W(\vr_2)\right].
\label{eq:muleak}
\ea
As the correlation function $\xi(\vr_1-\vr_2)$ strongly peaks at small $\vr_1-\vr_2$, 
we approximate that $W(\vr_1) \approx W(\vr_2)$. Since the $\mu$-leakage correction is already a second-order effect, we leave a rigorous study of the validity of this approximation to a future paper, and we find that for our purposes the resulting $\mu$-leakage matrix sufficiently captures the corrections from a naive summation over modes. 
Thus, we set $W(\vr_1)=W(\vr_2)$, which 
forces $\vq=\vk$. We now take the angular average by integrating over a thin shell of Fourier-space volume $V_\mathrm{shell}(q_i)$ around some $q_i$,
\ba
\label{eq:Mlm_continuous}
\int_{V_\mathrm{shell}(q_i)}\frac{\dd^3q}{V_\mathrm{shell}(q_i)}
\,\frac{F_L(\vq)\,F_0^*(\vq)}{V_\mathrm{survey}}
&=
\sum_{\ell}
P_{\ell}(q_i)
\left[
\int_{V_\mathrm{shell}(q_i)}\frac{\dd^3q}{V_\mathrm{shell}(q_i)}
(2L+1)\int\frac{\dd^3r_1}{V_\mathrm{survey}}
\,W^2(\vr_1)
\,\mathcal{L}_{L}(\qhat \cdot \rhat_1)
\,\mathcal{L}_{\ell}(\qhat \cdot \rhat_1)
\right],
\ea
where we use Eq.~\ref{eq:Fl} to identify the $F_L(\vq)$, and we assume that $P_\ell(q)\simeq P_\ell(q_i)$ is approximately constant across the shell around $q_i$. 

In identifying the $F_L(\vq)$ the Fourier transforms from $F(\vr)$ to $F_L(\vq)$ were discretized. Discretizing the other integrals in Eq.~\ref{eq:Mlm_continuous}, we get
\ba
\label{eq:pkl_Mlm}
\hat P_L(q_i)
&=
\sum_\ell \mathcal{M}_{L\ell}(q_i)\, P_\ell(q_i)\,,
\ea
where we define the $\mu$-leakage matrix as the term in brackets in Eq.~\ref{eq:Mlm_continuous}, with discretized integrals,
\ba
\label{eq:Mlm_win}
\mathcal{M}_{\ell L}(q_i)
&=
\frac{2L+1}{N_{\mathrm{modes}}(q_i)\,N_\mathrm{survey}}
\sum_{q_j\sim q_i}
\sum_{n}
\,W^2(\vr_{1,n})
\,\mathcal{L}_{L}(\qhat_j \cdot \rhat_{1,n})
\,\mathcal{L}_{\ell}(\qhat_j \cdot \rhat_{1,n})
\,,
\ea
where the first sum over $j$ is taking an average over all $N_{\mathrm{modes}}(q_i)$ modes within a thin shell around $q_i$, the sum over $n$ is taking an average over the survey in configuration space, and $N_\mathrm{survey}$ is the number of grid cells in the survey.
Eq.~\ref{eq:Mlm_win} is consistent with the case of no survey window (i.e $W$ identically equal to unity), and also with the result in the plane-parallel limit, Eq.~\ref{eq:Mlm_pp}.

To calculate the $\mu$-leakage matrix, the form Eq.~\ref{eq:Mlm_win} requires summing over $\sim N^6$ elements in total. We may simplify by splitting the Legendre polynomials into sums over spherical harmonics using Eq.~\ref{eq:legendre_ylmylm}, to obtain
\ba
\mathcal{M}_{\ell L}(q_i)
&=
\frac{(4\pi)^2}{2\ell+1}\sum_{Mm}
\left[
\frac{1}{N_{\mathrm{modes}}(q_i)}\sum_{q_j\sim q_i}
Y_{LM}(\qhat_j)
\,Y_{\ell m}(\qhat_j)
\right]
\left[
\frac{1}{N_\mathrm{survey}}\sum_{n}
\,W^2(\vr_{1,n})
\,Y^*_{LM}(\rhat_{1,n})
\,Y^*_{\ell m}(\rhat_{1,n})
\right]\,.
\ea
To reduce further we use Eq.~\ref{eq:YlmYlm},
\ba
\mathcal{M}_{\ell L}(q_i)
&=
\frac{(4\pi)^2}{2\ell+1}\sum_{Mm}
\frac{1}{N_{\mathrm{modes}}(q_i)}\sum_{q_j\sim q_i}
\frac{1}{N_\mathrm{survey}}\sum_{n}
\,W^2(\vr_{1,n})
\vs&\quad\times
\sum_{L'M'}
(-1)^{M'}
\sqrt{\frac{(2L+1)(2\ell+1)(2L'+1)}{4\pi}}
\begin{pmatrix}
  L & \ell & L' \\
  0 & 0  & 0
\end{pmatrix}
\begin{pmatrix}
  L & \ell & L' \\
  M & m & -M'
\end{pmatrix}
\,Y_{L'M'}(\qhat_j)
\vs&\quad\times
\sum_{L''M''}
(-1)^{M''}
\sqrt{\frac{(2L+1)(2\ell+1)(2L''+1)}{4\pi}}
\begin{pmatrix}
  L & \ell & L'' \\
  0 & 0  & 0
\end{pmatrix}
\begin{pmatrix}
  L & \ell & L'' \\
  M & m & -M''
\end{pmatrix}
\,Y^*_{L''M''}(\rhat_{1,n})
\,.
\ea
Using the orthogonality of the Wigner $3j$-symbols Eq.~\ref{eq:3j_orthogonality}, we get
\ba
\label{eq:Mlm_q}
\mathcal{M}_{\ell L}(q_i)
&=
4\pi\,(2L+1)
\sum_{L'M'}
\begin{pmatrix}
  L & \ell & L' \\
  0 & 0  & 0
\end{pmatrix}^2
\left[
\frac{1}{N_{\mathrm{modes}}(q_i)}\sum_{q_j\sim q_i}
\,Y_{L'M'}(\qhat_j)
\right]
\left[
\frac{1}{N_\mathrm{survey}}\sum_{n}
\,W^2(\vr_{1,n})
\,Y^*_{L'M'}(\rhat_{1,n})
\right]
\,.
\ea
The sum over $L'$ is from 0 to $2\ell_{\max}=8$.
Thus, the $\mu$-leakage matrix calculation is reduced to $2 \cdot 81 = 162$ sums over $N^3$ elements.

\subsection{\texorpdfstring{$\mu$}{Mu}-leakage discussion}
\label{sec:mu_discussion}
To facilitate the discussion of the $\mu$-leakage matrix \cref{eq:Mlm_q}, we write it as
\ba
\label{eq:Mlm_BLMq_CLM}
\mathcal{M}_{\ell L}(q_i)
&=
4\pi \, (2L+1)
\sum_{L'M'}
\begin{pmatrix}
  L & \ell & L' \\
  0 & 0  & 0
\end{pmatrix}^2
B_{L'M'}(q_i)
\,C_{L'M'}
\,,
\ea
where we defined
\ba
\label{eq:B_LM_q}
B_{L'M'}(q_i)
&=
\frac{1}{N_{\mathrm{modes}}(q_i)}\sum_{q_j\sim q_i}
\,Y_{L'M'}(\qhat_j)
\,,
\ea
and
\ba
\label{eq:C_LM}
C_{L'M'}
&=
\frac{1}{N_\mathrm{survey}}\sum_{n}
\,W^2(\vr_{1,n})
\,Y^*_{L'M'}(\rhat_{1,n})
\,.
\ea
If there is a large number of Fourier-space cells inside a shell of radius
$q_i$ that are evenly spaced on the sphere, then \cref{eq:B_LM_q} implies that
$B_{L'M'}(q_i)\approx\delta^K_{L'0}\delta^K_{M'0}/\sqrt{4\pi}$, which with the
Wigner-$3j$ symbol forces $\ell=L$ in \cref{eq:Mlm_BLMq_CLM}. Thus, compared to small scales, the
$\mu$-leakage correction is more important on large scales where there are few
modes to sum over and this approximation breaks down. Furthermore, because of
the parity of the spherical harmonics, $B_{L'M'}(q_i)=0$ for all odd $L'$.
Therefore, the Wigner-$3j$ forces $\Delta\ell=\ell-L$ to be even. Hence,
$\mu$-leakage only mixes even multipoles with even multipoles, and odd
multipoles with odd multipoles.

The $\mu$-leakage matrix depends on the survey geometry through $C_{L'M'}$. If
the survey is far from the observer (e.g., in the plane-parallel limit), then
the spherical harmonic in \cref{eq:C_LM} is constant and can be taken outside
the sum. Therefore, the $\mu$-leakage matrix can deviate significantly from
the unit matrix in this case. If, on the other hand, the window is a perfectly
spherically symmetric shell of some thickness $\Delta r$, then the sum in
\cref{eq:C_LM} is a good approximation to the integral, and we have
$C_{L'M'}\approx\delta^K_{L'0}\delta^K_{M'0}/\sqrt{4\pi}$. Because the number
of cells summed over is typically much greater here than the similar sum in
\cref{eq:B_LM_q}, this approximation is better than the one
for $B_{L'M'}$, especially for a thick shell. Therefore, if the window is
spherically symmetric around the observer, then the $\mu$-leakage matrix is
close to the unit matrix.

\begin{figure*}
  \centering
  \includegraphics[width=0.48\textwidth]{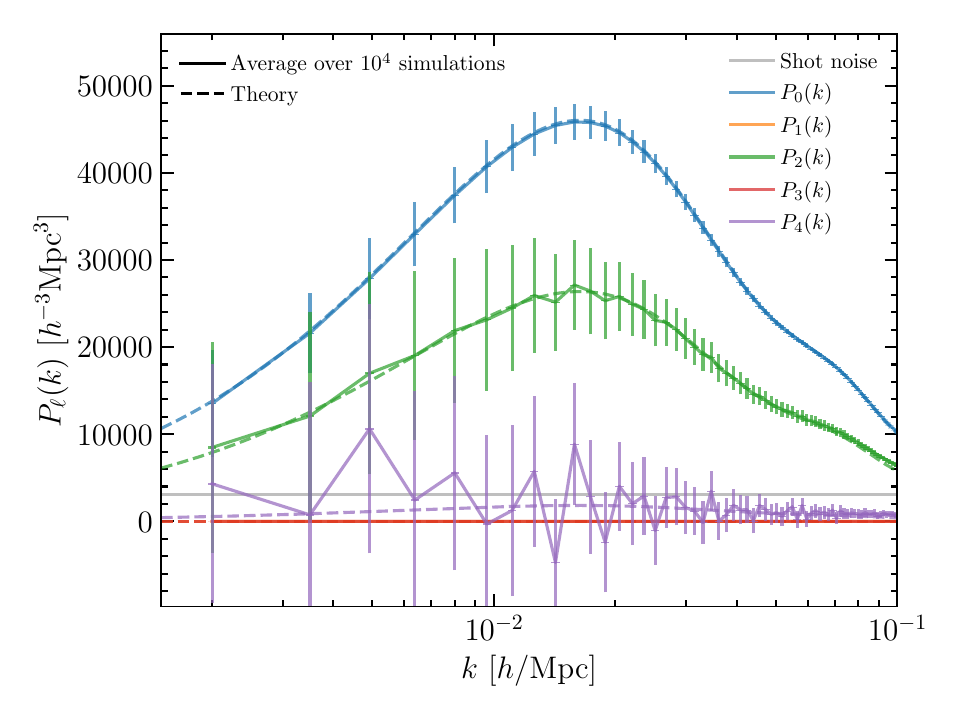}
  \includegraphics[width=0.48\textwidth]{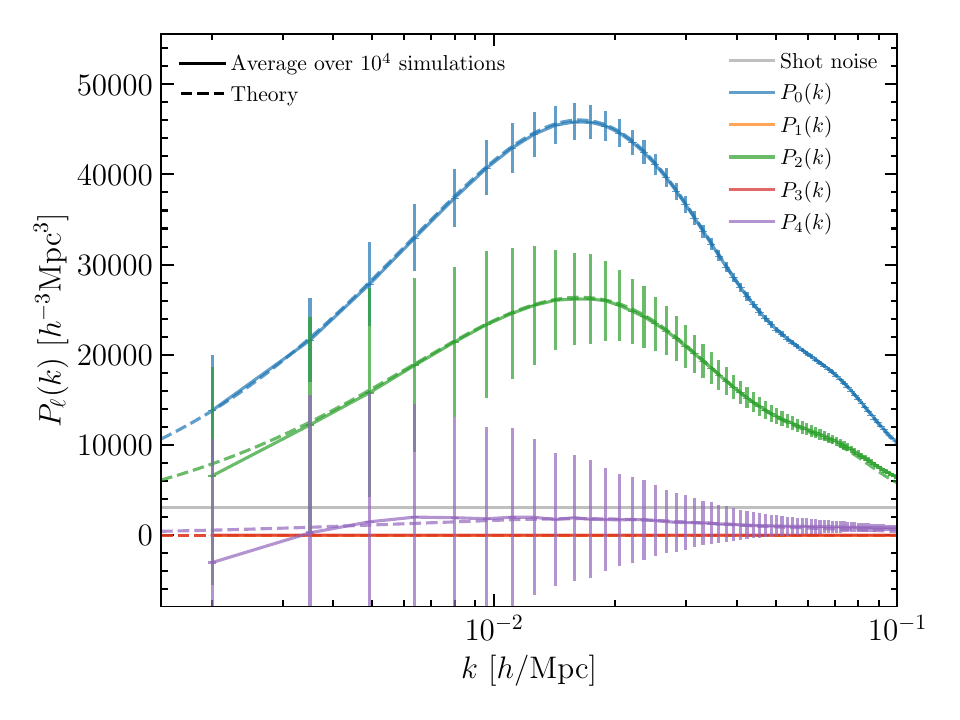}
  \caption{
    Here we show the power spectrum multipoles' averages over $10^4$ lognormal
    simulations with no window and periodic boundary conditions in the
    plane-parallel approximation (solid curves with
    error bars estimated from the simulations).
    The linear Kaiser theory is also shown
    (dashed curves). The left plot shows the simulation average without
    correcting for $\mu$-leakage, while the right plot uses our $\mu$-leakage
    correction.
  }
  \label{fig:pkl_pp_nowin}
\end{figure*}

We show the effect of the $\mu$-leakage in \cref{fig:pkl_pp_nowin}, where in
the left panel we show the power spectrum multipoles in the plane-parallel
approximation without $\mu$-leakage correction, and the right panel shows the
same but with $\mu$-leakage correction applied. The uncorrected quadrupole and
hexadecapole show spurious oscillations that largely disappear once the
correction is applied. The errorbars are at the location of the average
$k$-mode within each shell. We conclude from the right panel in the figure that the monopole is
accurate across the plot; the quadrupole is inaccurate only at the very
largest mode and the very smallest modes; and the hexadecapole has significant
inaccuracies across multiple modes. Thus, we caution the use of the
hexadecapole from the lognormal simulations for comparison on large scales.

\end{document}